\documentclass[%
 reprint,
 superscriptaddress,
 amsmath,amssymb,
 aps,
floatfix,
]{revtex4-2}

\usepackage{color}

\usepackage{graphicx}% Include figure files
\usepackage{dcolumn}% Align table columns on decimal point
\usepackage{bm}% bold math
\usepackage{hyperref}% add hypertext capabilities
\usepackage[dvipsnames]{xcolor} %Allows different textcolors
\usepackage{ulem}  %Allows strikethrough
\begin{document}

%\preprint{APS/123-QED}

\title{Perpendicular and Parallel Phase Separation in Two Species Driven Diffusive Lattice Gases}

\author{Honghao Yu}
\affiliation{%
 Yusuf Hamied Department of Chemistry, University of Cambridge, Lensfield Road, Cambridge CB2 1EW, United Kingdom}%
\author{Kristian Thijssen}%
\affiliation{%
 Yusuf Hamied Department of Chemistry, University of Cambridge, Lensfield Road, Cambridge CB2 1EW, United Kingdom}%
\author{Robert L. Jack}%
\affiliation{%
 Yusuf Hamied Department of Chemistry, University of Cambridge, Lensfield Road, Cambridge CB2 1EW, United Kingdom}%
\affiliation{%
 Department of Applied Mathematics and Theoretical Physics, University of Cambridge, Wilberforce Road,
Cambridge CB3 0WA, United Kingdom}%

\newcommand{\beq}{\begin{equation}}
\newcommand{\eeq}{\end{equation}}
\newcommand{\CorrectText}[2]{{\sout{#1}}{~{#2}}}
\normalem % ulem makes emph into an underline, so this fixes that

\date{\today}% It is always \today, today,

\begin{abstract}
We study three different lattice models in which two species of diffusing particles are driven in opposite directions by an electric field.  
We focus on dynamical phase transitions that involve phase separation into domains that may be parallel or perpendicular to a driving field. 
In all cases, the perpendicular state appears for weak driving, consistent with previous work.  For strong driving, we introduce two models that support the parallel state.  In one model, this state occurs because of the inclusion of dynamical rules that enhance lateral diffusion during collisions; in the other, it is a result of a nearest-neighbour attractive/repulsive interaction between particles of the same/opposite species.  We discuss the connections between these results and the behaviour found in off-lattice systems, including laning and freezing by heating.
\end{abstract}

\maketitle

%\tableofcontents

\section{\label{sec:introduction}Introduction}

Non-equilibrium systems exhibit a wide variety of fascinating phenomena. Examples are systems consisting of self-propelled particles such as bacteria \cite{thompson2011lattice,fily2012athermal}, flocks of birds \cite{bialek2012statistical} and even human crowds \cite{henderson1971statistics}. Another class of non-equilibrium systems are systems driven away from equilibrium by external energy sources such as electric fields \cite{katz1984nonequilibrium} or temperature gradients \cite{duhr2006molecules}. Both types of non-equilibrium systems possess non-equilibrium steady states, distinguished from equilibrium by breaking of detailed balance and time-reversal symmetry~\cite{seifert2012stochastic, chou2011non, marchetti2013hydrodynamics}. 

Well-studied examples of such states include mixtures of two species, driven in opposite directions, in two dimensions. These might consist of passive particles like colloids~\cite{leunissen2005ionic, vissers2011lane, vissers2011band, helbing2000freezing, dzubiella2002lane, glanz2012nature, wachtler2016lane, klymko2016microscopic, dutta2016anomalous, reichhardt2018velocity, dutta2018transient, dutta2020length, geigenfeind2020superadiabatic, li2021phase}, or active agents like humans or ants~\cite{helbing1995social, couzin2003self, karamouzas2014universal, oliveira2016keep, reichhardt2018laning}.  Under such conditions, the driven particles or active agents may follow each other, avoiding collision with  oppositely moving particles, which is called \emph{laning}~\cite{vissers2011lane}.  Alternatively, the driving may cause particles to block each other, similar to a \emph{traffic jam}~\cite{helbing2000freezing} or -- for alternating fields -- one may observe moving \emph{bands} with high-density \cite{vissers2011band}.  In experiments of two-component colloidal mixtures, both lane formation~\cite{vissers2011lane} and band formation \cite{vissers2011band} are observed.

To understand the laning phenomenon from a theoretical perspective, Brownian dynamics simulations have been widely employed to study its properties~\cite{helbing2000freezing, dzubiella2002lane, chakrabarti2003dynamical, chakrabarti2004reentrance, glanz2012nature, kohl2012microscopic, wachtler2016lane, klymko2016microscopic, oliveira2016keep, poncet2017universal, ikeda2017lane, reichhardt2018velocity, reichhardt2018laning, liu2019controllable, li2021phase}.  While the simplest models only support laning~\cite{dzubiella2002lane, chakrabarti2003dynamical, lowen2003nonequilibrium, chakrabarti2004reentrance, glanz2012nature, kohl2012microscopic, klymko2016microscopic, poncet2017universal}, they can be modified to support traffic jam behaviour, for example, by removing thermal noise~\cite{reichhardt2018velocity, reichhardt2018laning} or using long-ranged repulsive~\cite{helbing2000freezing} or short-range attractive potentials~\cite{wachtler2016lane}.

Dynamical phase transitions and spontaneous symmetry breaking have been found in some of these previous investigated systems~\cite{helbing2000freezing, wachtler2016lane, reichhardt2018velocity, reichhardt2018laning}.  By analogy with equilibrium systems, a natural hypothesis is that such transitions might be described by universal theories of Landau-Ginzburg type. The analysis of simple (lattice-based) models should then provide insight (and quantitative predictions) for more complex off-lattice systems.  Among such lattice models, a driven diffusive lattice gas with a single particle species was proposed by Katz, Lebowitz, and Spohn (KLS)~\cite{katz1984nonequilibrium}: it exhibits phase separation into domains oriented parallel to the driving field.  This work was then broadened to two species driven in opposite directions where phase separation perpendicular to the driving field is supported~\cite{schmittmann1992onset, bassler1993spatial, foster1994finite, vilfan1994spontaneous, korniss1995novel, korniss1997nonequilibrium, korniss1999long}, see \cite{schmittmann1998driven} for a review.

\begin{figure}
\includegraphics[width=0.48\textwidth]{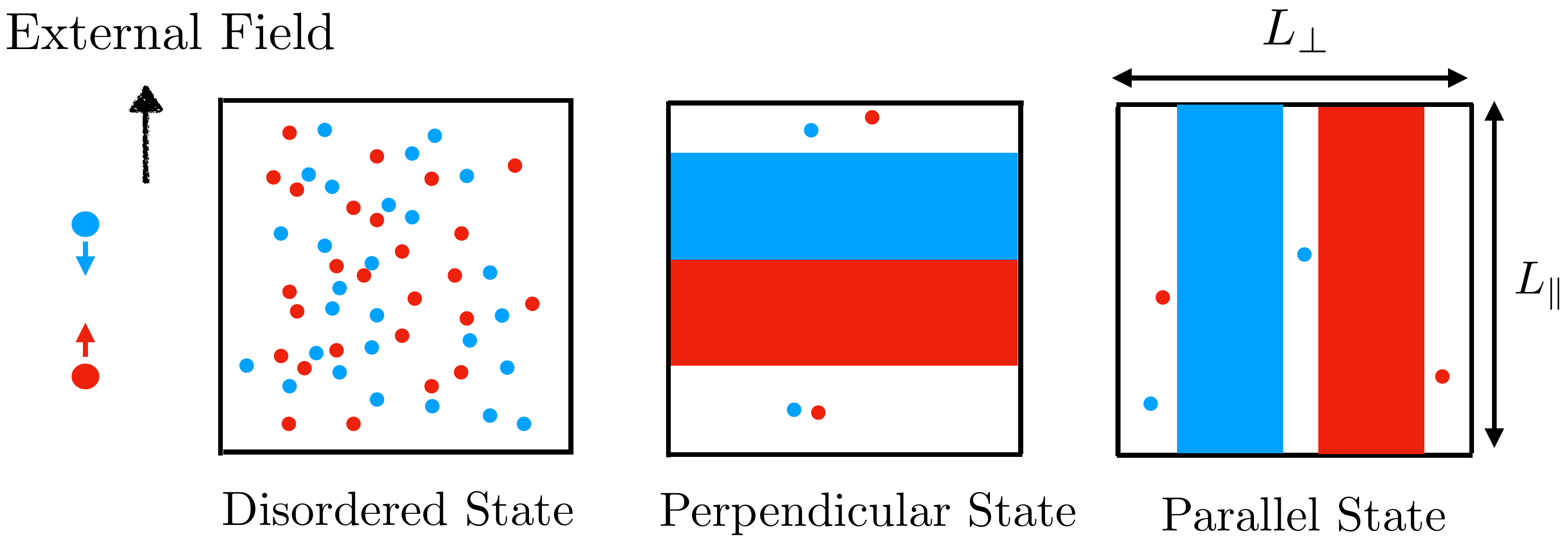}
\caption{\label{fig:SchematicPT} Three possible steady states for oppositely driven colloidal particles in two dimensions: disordered state, phase separation perpendicular to the external field, phase separation parallel to the external field. Positive particles (moving along the external field)
are labelled red and negative particles 
(moving opposite to the external field) are labelled blue.}
\end{figure}

Depending on the situation, two-species off-lattice and on-lattice models may support phase separation with domains parallel to the field, or perpendicular to it (see the schematic picture in Fig.~\ref{fig:SchematicPT}).  The parallel state is similar to laning (it corresponds to two macroscopic lanes) while the perpendicular state resembles a traffic jam, where the two species tend to block each other.  Note that if lanes form with a finite width -- as is often observed in experiments and simulations -- this emergence does not require a phase transition or spontaneous symmetry breaking.  However, it is often difficult to establish whether these finite lanes will persist as a non-equilibrium steady state, or if they should eventually coarsen into macroscopic lanes, see for example~\cite{klymko2016microscopic}.

Among the open questions in this area, there remain some differences in the reported behaviour of lattice and off-lattice models.  For example, lattice models that support the perpendicular state of Fig.~\ref{fig:SchematicPT}(b) were analysed already in the 1990s~\cite{schmittmann1992onset, bassler1993spatial, foster1994finite, vilfan1994spontaneous, korniss1995novel, korniss1997nonequilibrium, korniss1999long}, but these models do not support the parallel state observed in Brownian dynamics.
On the other hand, for Brownian dynamics simulations of oppositely driven particles with short-range repulsive interactions, only the parallel state is observed~\cite{dzubiella2002lane, lowen2003nonequilibrium, chakrabarti2003dynamical, chakrabarti2004reentrance, glanz2012nature, kohl2012microscopic, klymko2016microscopic, oliveira2016keep, poncet2017universal}, and the perpendicular state is absent.  It is also debated whether the transition to the parallel state is a genuine phase transition~\cite{glanz2012nature, kohl2012microscopic, klymko2016microscopic, poncet2017universal}.

To better understand these differences in behaviour between on-lattice and off-lattice models, this work analyses three different lattice models that include two species of oppositely driven (oppositely-charged) particles.  The first model is that of Schmittmann, Hwang, and Zia (SHZ)~\cite{schmittmann1992onset}, which supports disordered (homogeneous) and perpendicular states.  The other two models [an enhanced lateral diffusion (ELD) model~\cite{klymko2016microscopic} and a modified two-species Katz, Lebowitz and Spohn (TKLS) model] support both parallel and perpendicular states.  This shows that the parallel state~\cite{dzubiella2002lane,  klymko2016microscopic} can be observed in lattice systems, although extra assumptions are required when constructing such models. 

We find that the parallel state is stablilised by a combination of the driving field and an effective attraction/repulsion between particles of the same/opposite charge.  
In the enhanced lateral diffusion (ELD) model, the effective interactions are generated by an enhanced diffusivity in the direction perpendicular to the field, which occurs when oppositely moving particles collide.  This mechanism was suggested in~\cite{klymko2016microscopic}, see also~\cite{burger2016lane} for a setting with asymmetric lateral mobility (``side-stepping'').  
Our implementation of ELD shares some features with motility induced phase separation (MIPS) in active matter~\cite{cates2015motility}.  
In addition, we consider  a two-species generalisation of the KLS model (TKLS), in which effective (nearest-neighbour) interactions between like and unlike charges are included explicitly in the interaction energy.  
The  non-equilibrium steady states in these two models indicate that the parallel phase is robust in systems with the appropriate, effective interactions.

In addition to identifying the phases that appear in these models, we analyse the transitions between them. This will include a discussion of finite-size scaling, meta-stability, and the distributions of appropriate order parameters.  
We find a rich phenomenology including both continuous and discontinuous (first-order) transitions.
In addition, the transition between the parallel and perpendicular states in the ELD model appears to take place by an intermediate phase with an unusual zig-zag order.

The paper is organised as follows.
Sec.~\ref{sec:models} gives an overview of the models that we consider, and the order parameters that we will use to analyse the phase transitions.
Sec.~\ref{sec:shz} analyses the SHZ model, building on previous work  \cite{schmittmann1992onset, vilfan1994spontaneous, korniss1995novel, korniss1997nonequilibrium}, with an expanded numerical characterisation of the phase transition itself.  
Then, Sec.~\ref{sec:eld} and Sec.~\ref{sec:tkls} introduce and analyse the ELD and TKLS models, respectively.
Finally, Sec.~\ref{sec:conclusion} summarises the main insights of the work and the comparison with on- and off-lattice systems.

\section{Models and Order Parameters}
\label{sec:models}

\subsection{General definitions}

The models that we consider are two-species driven diffusive lattice gases.
They consist of  $N$ hard particles on a rectangular lattice of size $L_{\parallel}\times L_{\perp}$ with periodic boundary conditions.  All numerical results in this work are obtained for square lattices, $L_\perp = L_\parallel = L$,
but we maintain the possibility of other aspect ratios in our theoretical discussion. 

A constant external electric field $\bm{E}=E\hat{\bm{y}}$ is applied in the $+y$ direction driving the two species in opposite direction.  
The two species are indicated by the signs of their charges, $+$ and $-$, and are indicated by red and blue in Fig.~\ref{fig:SchematicPT} respectively.
The notation $\bm{x}=(x,y)$ indicates a lattice site and we define occupation variables $n^{+}_{\bm{x}}$ and $n^{-}_{\bm{x}}$ for the two species. 
The charge on site $\bm{x} $ is therefore 
\beq
\sigma_{\bm{x}} = n^{+}_{\bm{x}} - n^{-}_{\bm{x}}.
\eeq
Each site can be occupied by at most one particle so $n^\pm_{\bm{x}}$ has possible values $0,1$ and $\sigma_{\bm{x}}$ has possible values $0,\pm1$.   
Hence, it is natural to identify the lattice spacing with the particle diameter.

Particles can move by two processes: either a particle on site $\bm{x}$ can hop to a neighbouring vacant site $\bm{x}+\bm{e}$ with rate $W_{\rm H}(\bm{x},\bm{e},\sigma_{\bm{x}})$; or if two neighbouring sites $\bm{x},\bm{x}+\bm{e}$ are occupied by particles of different species, then these particles can swap places with rate $W_{\rm S}(\bm{x},\bm{e},\sigma_{\bm{x}})$.  Since all processes involve neighbouring sites then $|\bm{e}|=1$ in these rates.  The rates may also depend implicitly on the local environment of site $\bm{x}$ and on the field $E$.
For $E=0$, all hop rates are the same $W_{\rm H}=1$, which fixes the unit of time.
We implement these dynamics using a Monte Carlo (MC) method, details are given in Appendix~\ref{sec:app-mc}.

The number of particles of each species is conserved.
The total number of particles is $N=\sum_{\bm{x}} (n^{+}_{\bm{x}} + n^{-}_{\bm{x}})$ and the overall density is $\rho=N/(L_{\parallel} L_{\perp})$.
We consider systems at charge neutrality:  $\sum_{\bm{x}} \sigma_{\bm{x}}=0$.

\subsection{Order Parameters and spontaneous symmetry breaking}

Throughout this manuscript, we will characterise non-equilibrium phase transitions. While these transitions share many features with their equilibrium counterparts, they are not characterised by an underlying Boltzmann distribution or free energy.  Instead, we characterise phase transitions in terms of spontaneous breaking of symmetry: in this case, translational symmetry of the lattice along either the $x$ or $y$ direction, recall Fig.~\ref{fig:SchematicPT}.

To characterise this, we follow \cite{korniss1995novel,korniss1997nonequilibrium} and define Fourier transformed densities:
\begin{align}\label{eq:DefComplexOrder}
\phi({\bm{k}})&=\frac{1}{L_{\perp}L_{\parallel}}\sum_{\bm{x}} (1-n_{\bm{x}}^+ - n_{\bm{x}}^-) e^{-i\bm{k}\cdot \bm{x}},  \\
\psi({\bm{k}})&=\frac{1}{L_{\perp}L_{\parallel}}\sum_{\bm{x}} \sigma_{\bm{x}} e^{-i\bm{k}\cdot \bm{x}},
\end{align}
with $\bm{k}=2\pi(n_{\perp}/L_{\perp}, n_{\parallel}/L_{\parallel})$, for integers $n_{\perp}, n_{\parallel}$.  Physically, $\phi$ is the Fourier transform of the density of vacant sites and $\psi$ is the Fourier transform of the charge density, with complex magnitudes
\begin{equation}
\Phi({\bm{k}}) = |\phi(\bm{k})| , \qquad
\Psi({\bm{k}}) = |\psi(\bm{k})| . 
\end{equation}
The key point is that $\langle \phi({\bm{k}}) \rangle=0$ in any translationally invariant system, and $\langle \Phi({\bm{k}}) \rangle=O(1/L)$ in large systems. However, these values become non-trivial if symmetry is spontaneously broken.

To see this, consider the configurations sketched in Fig.~\ref{fig:SchematicPT}(b,c) where the formation of high-density regions breaks translational symmetry.  To distinguish the situations shown in that Figure, we consider two specific wavevectors
\beq
\bm{k}_\parallel = (2\pi/L_\perp,0), \qquad \bm{k}_\perp = (0,2\pi/L_\parallel). 
\eeq
These are the smallest accessible wavevectors in this system, corresponding to $(n_\perp,n_\parallel)=(1,0)$ or $(0,1)$.
The notation is chosen because the order parameter $\phi(\bm{k}_\parallel)$ now distinguishes 
the parallel phase [Fig.~\ref{fig:SchematicPT}(c)].  In these configurations, $\phi(\bm{k}_\parallel)$ is a complex number whose argument depends on the position of the dense region (along the $y$-axis) and whose modulus measures the difference in density between the dense and dilute regions.   

If the density difference between dense and dilute regions remains of order unity as the system size increases, then the modulus of $\phi(\bm{k}_\parallel)$ is also of order unity, but its argument is random (because the position of the dense region is random).  Hence, taking an ensemble average, $\langle \phi(\bm{k}_\parallel)\rangle=0$ but 
$\langle \Phi(\bm{k}_\parallel)\rangle=O(1)$.  On the other hand, for homogeneous phases $\langle \Phi({\bm{k}_\parallel}) \rangle=O(1/L)$.

This observation allows a precise definition of dynamical phase transitions: one defines an order parameter as
$
\Phi_{\parallel,\infty} = \lim_{L\to\infty} \langle \Phi(\bm{k}_\parallel)\rangle,
$
which is positive in the ordered (symmetry-broken) phase and zero in disordered (homogeneous) systems.
We determine the behaviour of the order parameters with finite-size scaling analysis throughout this manuscript.

We simulate finite systems with fixed particle numbers, and the dynamical rules mean that every configuration is accessible from every other.  Hence the systems are ergodic.  Averages are computed by extracting many configurations from long dynamical trajectories. The early parts of the trajectories are discarded, to ensure that the systems have converged to their steady states.  (See also the discussion of Fig.~\ref{fig:FORealOPSHZMain}, below.)
 
For the ordered phase shown in Fig.~\ref{fig:SchematicPT}(b), a similar argument holds with $\bm{k}_\perp$ instead of $\bm{k}_\parallel$.  Similar arguments also hold if the ``density'' order parameter $\phi$ is replaced by the ``charge'' order parameter $\psi$.   Whether we use $\phi$ or $\psi$ as order parameter depends on the detailed behaviour of the inhomogeneous states, which will be explained in the relevant sections. 
 
\subsection{Density and charge profiles for inhomogeneous states}
\label{sec:recen}

For states where the symmetry is broken, we compute the associated density and charge profiles.  In long simulations, the regions of high or low density can diffuse, so we recenter the system at each time, to measure a meaningful profile, following previous work~\cite{krometis2004lane}.   Given a broken symmetry state with characteristic wavevector $\bm{k}$, define $\theta = \operatorname{arg}(\phi(\bm{k}))$.  For the perpendicular state one takes $\bm{k}=\bm{k}_\perp$, this means that the center of the dilute region of the system has position close to $(L_\perp/2,y^*)$ with $y^* = L\theta/(2\pi)$.   We then compute the average density as a function of the vertical co-ordinate $y-y^*$, which is measured relative to the center of the dilute region.  [See for example Fig.~\ref{fig:ConfigurationSHZ}(b) below.]  A similar procedure is used for the parallel state, see for example Fig.~\ref{fig:ConfigurationTKLS}.

\section{SHZ model : continuous and discontinuous phase transitions}
\label{sec:shz}

\subsection{Model definition}

In the SHZ model, particle hop rates are determined by a Metropolis formula \cite{metropolis1953equation}.  Note that $\sigma_{\bm{x}}\bm{E}\cdot\bm{e}$ is the work done by the electric field for a hop along vector $\bm{e}$ by a particle of species $\sigma_{\bm{x}}$.  There are no energetic interactions between the particles, and the rates $W_{\rm H},W_{\rm S}$ depend only on $(\sigma_{\bm{x}},\bm{e})$.
Specifically, we take
\begin{equation}\label{eq:SHZRegularHopRate}
W_{\text{SHZ,H}}(\bm{x},\bm{e},\sigma_{\bm{x}}) = \min(1, \exp(\sigma_{\boldsymbol{x}}\bm{E}\cdot\bm{e})).
\end{equation}
The corresponding swap rate is similar
\begin{equation}\label{eq:SHZRegularSwapRate}
W_{\text{SHZ,S}}(\bm{x},\bm{e},\sigma_{\bm{x}}) = \gamma\min(1, \exp(2\sigma_{\boldsymbol{x}}\bm{E}\cdot\bm{e})), 
\end{equation}
where $\gamma$ is a parameter that controls the relative rate of hops and swaps, and the factor of 2 in the exponent appears because two oppositely charged particles are moving in opposite directions, which doubles the work done.   We take $\gamma<1$, the physical role of these swaps is discussed below.  Note that this $\gamma$ differs from that of~\cite{korniss1995novel, korniss1997nonequilibrium}  by a factor of 2.

The combination of the dynamical rules \eqref{eq:SHZRegularHopRate} and \eqref{eq:SHZRegularSwapRate} with the periodic boundaries mean that for $\bm{E}\neq0$, the steady state of this system supports particle currents, the system is far from equilibrium, and the dynamics do not obey the principle of (global) detailed balance~\cite{schmittmann1998driven}.  However, it is useful to note that
the model is consistent with a (weaker) principle of \emph{local} detailed balance at temperature $T=1$, that is
\beq
\frac{ W_{\text{SHZ,H}}(\bm{x},\bm{e},\sigma_{\bm{x}}) }{ W_{\text{SHZ,H}}(\bm{x}+\bm{e},-\bm{e},\sigma_{\bm{x}})) } = \exp(\sigma_{\bm{x}}\bm{E}\cdot\bm{e}) \;.
\label{equ:SHZ-detbal}
\eeq
A similar relationship holds for swap moves.  
To see the relevance of this condition: note that if the periodic boundaries of the system are replaced by closed boundaries (hard walls), then the field $\bm{E}$ can be written as a the gradient of a potential, and (\ref{equ:SHZ-detbal}) ensures global detailed balance, leading to an equilibrium (Boltzmann-like) steady state without any currents.  In this sense, the local hopping rules (\ref{eq:SHZRegularHopRate}, \ref{eq:SHZRegularSwapRate}) are consistent with the equilibrium dynamics of particles in an electric field.  If the principle of local detailed balance is broken (as will happen later in the ELD model), the system can no longer be transformed into an equilibrium system by changing the boundary conditions.

\begin{figure}
    \centering
    \includegraphics[width=0.48\textwidth]{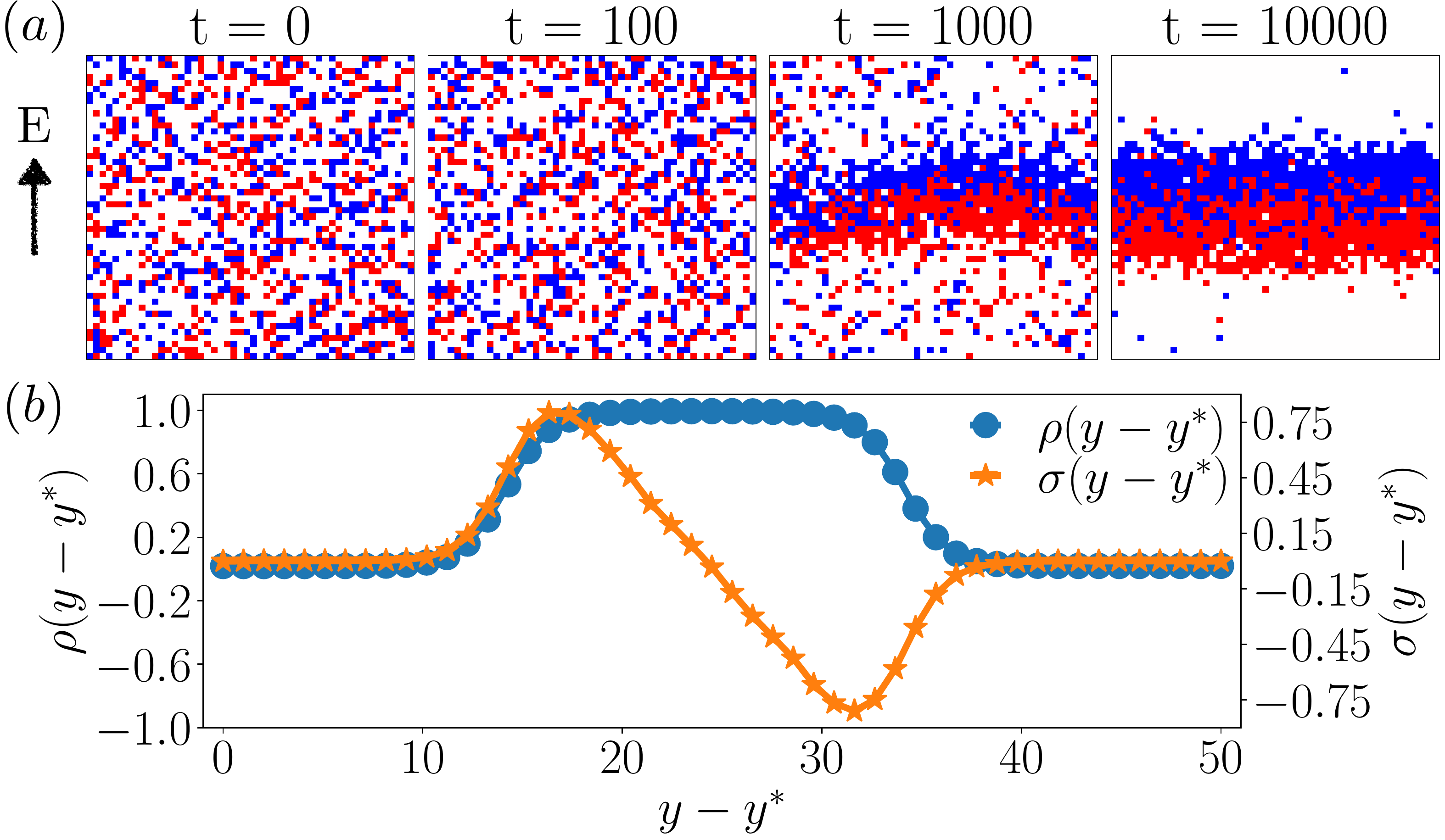}
    \caption{(a). Time series of configurations from a trajectory of the SHZ model with $E=1$, $\rho=0.4$ and $\gamma=0.01$ on a $50\times 50$ lattice. The model exhibits a phase transition between the disordered phase and perpendicular phase.
    (b). Steady-state particle density $\rho(y-y^*)$ and charge density $\sigma(y-y^*)$ profile along the driven direction, for the same control parameters as (a).  The average is computed from a single long trajectory after the system has reached its steady state, the averaging time is $\tau = 4\times 10^4$. Vacancy density $\phi(\bm{k})$ is used to recenter the density profiles (See Sec. \ref{sec:recen}.).
    }
    \label{fig:ConfigurationSHZ}
\end{figure}

Details of the implementation of the SHZ model are given in Appendix~\ref{sec:app-shz}. 

\subsection{Phase diagram}\label{sec:shz-phase}
For sufficiently large fields $E$ and densities $\rho$, the SHZ model exhibits spontaneous symmetry breaking [Fig.~\ref{fig:ConfigurationSHZ}(a)] and forms the perpendicular state shown in Fig.~\ref{fig:SchematicPT}(b).  This phase transition has been extensively studied by Korniss et. al. \cite{korniss1995novel, korniss1997nonequilibrium}.  In this section, we illustrate the main features of this phase transition with numerical simulations, similar to those of~\cite{korniss1997nonequilibrium}.  
In particular, we demonstrate the finite-size scaling behaviour of the order parameters discussed above.  This gives us a baseline to compare the results of our other two models with.  As usual, phase transitions are well-defined in large systems: we consider finite-size scaling where both $L_\parallel$ and $L_\perp$ are proportional to a finite-size scaling length parameter $L$. 

The physical origin of the perpendicular state [Fig.~\ref{fig:SchematicPT}(b)] is that oppositely charged particles are driven in  opposite directions and tend to impede each other.  In particular, if a negative particle blocks a positively-charged one, then other positive particles will tend to queue up behind the blocked particle.  If this effect is strong enough, a macroscopic traffic jam can form.  Such effects occur in various systems \cite{Biham1992Self, chowdhury2000statistical,arndt1998spontaneous, arndt1999spontaneous, clincy2003phase, kourbane2018exact}.  

The structure of this traffic jam can be observed from the steady-state density profiles of particles and charges as in Fig.~\ref{fig:ConfigurationSHZ}(b), whose computation was described in Sec.~\ref{sec:recen}.  It is notable that the charge profile varies smoothly across the dense region. There is no clearly-defined interface between red and blue domains.  The interfaces between dense and dilute regions are more clearly defined.

\begin{figure}
    \centering
    \includegraphics[width=0.48\textwidth]{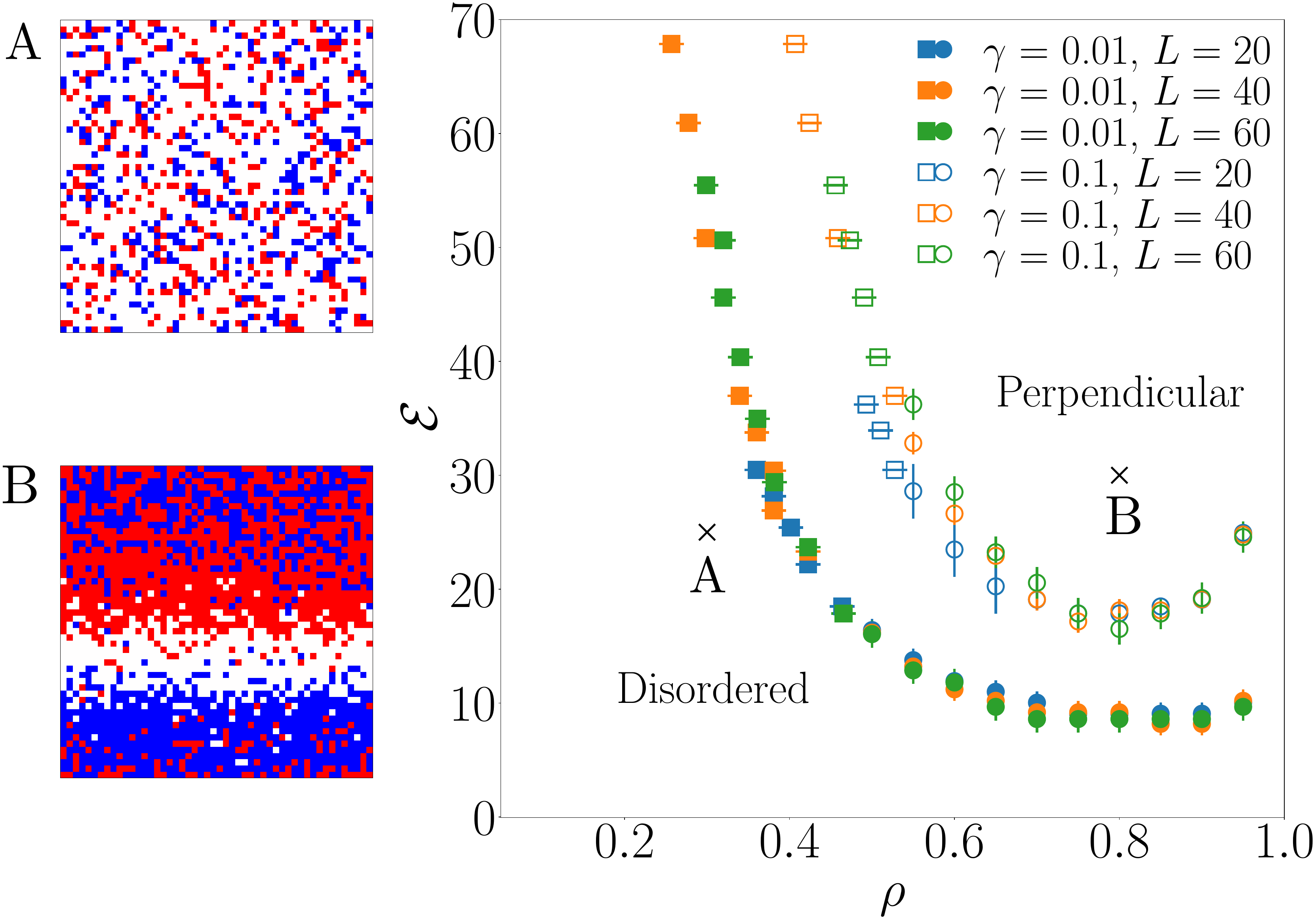}
    \caption{
    Phase Diagram of the SHZ model for various system sizes, with $\gamma = 0.01$, $0.1$. The system consists of a disordered phase and a perpendicular phase separated by a discontinuous (square) or continuous (circle) transition.}
    \label{fig:PhaseDiagramSHZ}
\end{figure}

As shown in~\cite{korniss1997nonequilibrium}, it is natural to reparameterise the dependence on $E$ when discussing the disorder-perpendicular transition in terms of 
\beq
{\cal E} = 2 L_{\parallel} \tanh{(E/2)} .
\label{equ:calE}
\eeq
Note in particular that for large $L_\parallel$, a finite value of ${\cal E}$ corresponds to a very small value of $E$ (which is $O(1/L_\parallel)$). 
The size dependency in the transition can be rationalised by a mean field analysis \cite{vilfan1994spontaneous, korniss1995novel, korniss1997nonequilibrium} and is also observed in other driven diffusive systems \cite{clincy2003phase, kourbane2018exact}.

\begin{figure*}
    \centering
    \includegraphics[width=0.87\textwidth]{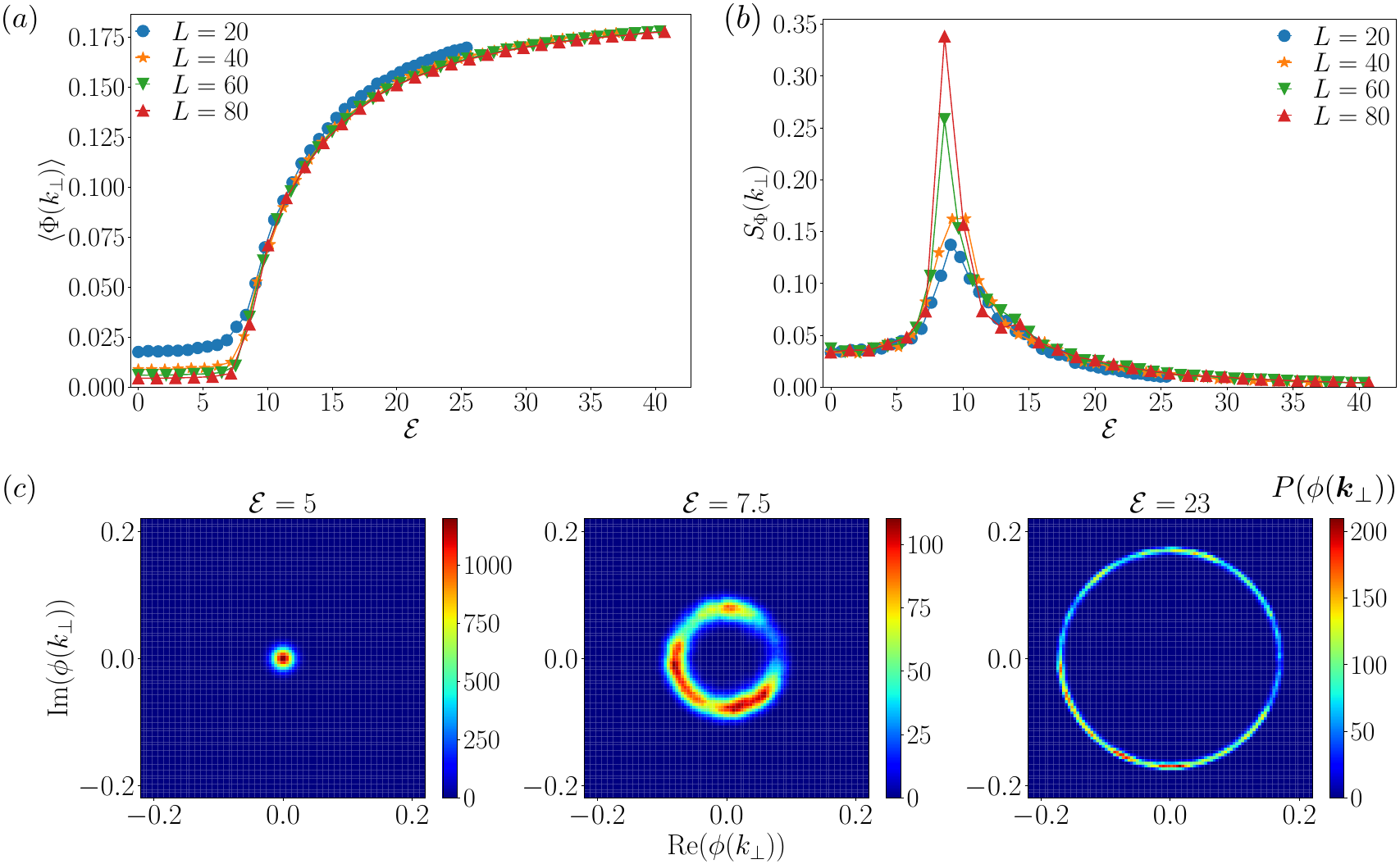}
    \caption{(a)~Order parameter $\Phi(\bm{k}_\perp)$ vs $\cal E$. (b)~Scaled variance $S_{\Phi}(\bm{k}_\perp$) vs $\cal E$ for the continuous transition with $\rho=0.8$, $\gamma=0.01$ for four different system sizes in the SHZ model. 
    (c)~The probability distribution of the complex order parameter $P(\phi(\bm{k}_\perp))$ across the continuous transition region for $\rho=0.8$, $\gamma=0.01$, and $L_{\parallel}=L_{\perp}=20$.}%
    \label{fig:SORealOPSHZMain}%
\end{figure*}

\begin{figure*}
    \centering
    \includegraphics[width=0.87\textwidth]{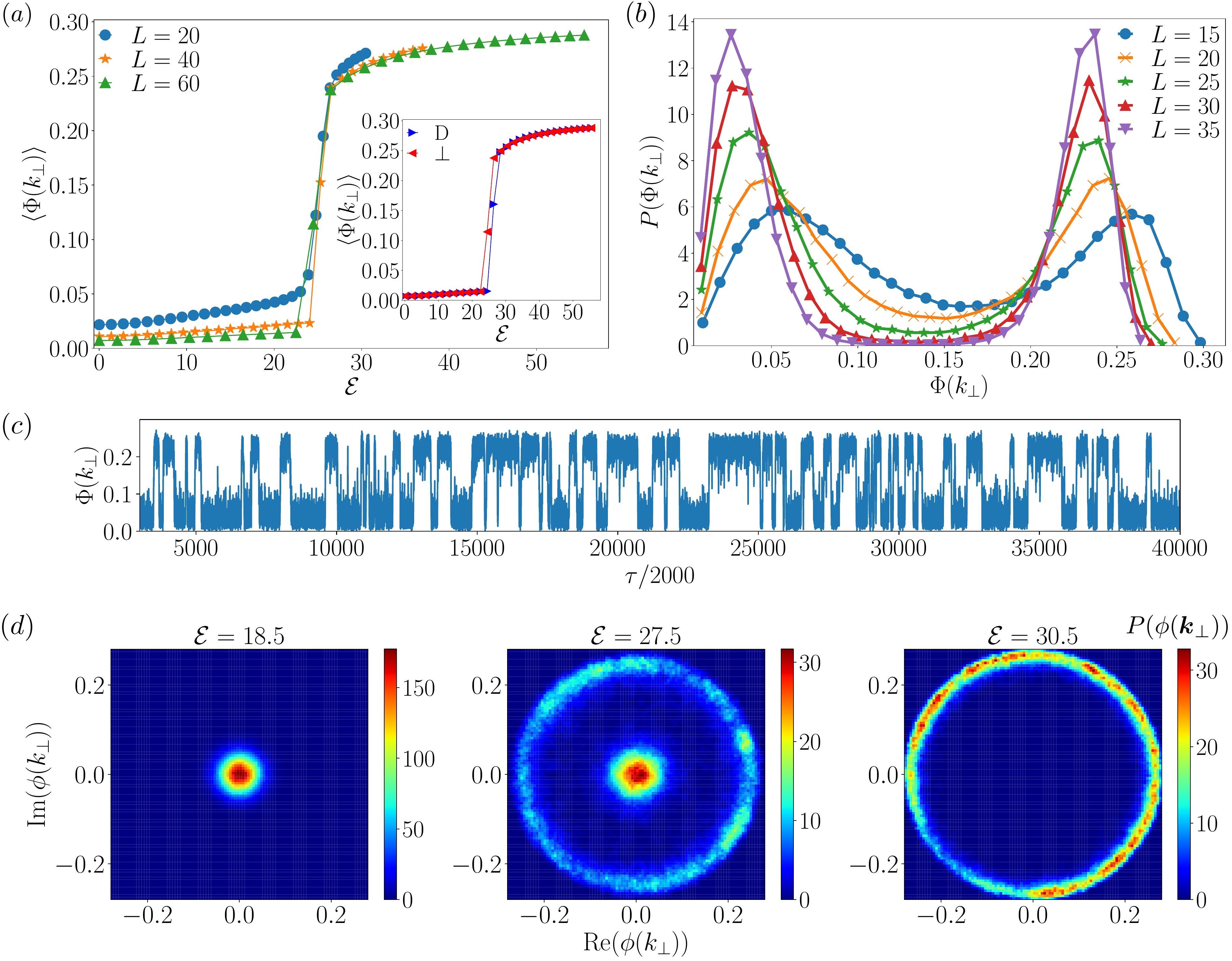}
    \caption{(a)~Order parameter $\Phi(\bm{k}_\perp)$ vs $\cal E$ for the discontinuous transition with $\rho=0.4$, $\gamma=0.01$ for four different system sizes in the SHZ model. 
    The inset shows that the results do not depend on the initial condition:  results are shown for simulations starting in disordered (D) or ordered perpendicular states ($\perp$), which show almost identical behavior.  Each point is an independent simulation. Data are shown for $L = 60$.
    (b)~The probability distribution of order parameter $P(\Phi(\bm{k}_\perp)$) at five different system sizes with $\rho=0.4$, $\gamma=0.01$ and $\cal E\approx$ $25$ 
    (c)~Time series of the order parameter $\Phi(\bm{k}_\perp)$ corresponding to panel (b), for system size $L_{\parallel}=L_{\perp}=25$. (d)~The probability distribution of the complex order parameter $P(\phi(\bm{k}_\perp)$) across the discontinuous transition region for $\rho=0.4$, $\gamma=0.01$, and $L_{\parallel}=L_{\perp}=25$.}%
    \label{fig:FORealOPSHZMain}%
\end{figure*}

Fig.~\ref{fig:PhaseDiagramSHZ} shows a dynamical phase diagram of this model, as a function of $\rho,{\cal E}$.  
It shows disordered (homogeneous) and perpendicular states, separated by a transition at ${\cal E}={\cal E}_c>0$.
The estimates of the phase boundaries in Fig.~\ref{fig:PhaseDiagramSHZ} are based on maximisation of the variance of the order parameter; see below for further details.   We run long simulations up to total time $\tau = 6\times 10^8$ and discard the data for time $\tau<4\times 10^7$ for each set of control parameters.  
We checked that this time is long enough for the systems to converge to their steady states.

The phase diagram shows that the system is in the perpendicular state for ${\cal E}>{\cal E}_c(\rho,\gamma)$, and disordered for smaller ${\cal E}$. The ${\cal E}$ factor provides good data collapse for systems with different system sizes.
However, the collapse is not so good for systems with $(L,\gamma)=(20,0.1)$: this is due to small system sizes, together with the fact that one always has ${\cal E}\leq 2L_\parallel$ [due to \eqref{equ:calE}].
The transition between disordered and perpendicular states may be continuous or discontinuous, as already explained theoretically in~\cite{korniss1997nonequilibrium}; similar behaviour is also observed in one dimension \cite{clincy2003phase}.  Specifically, the phase transition is continuous for large $\rho$, and discontinuous for smaller $\rho$.
The change between these two behaviours occurs at a $\gamma$-dependent value of $\rho$.

Before analysing the phase transitions, we briefly discuss the role of $\gamma$.  It is clear from Fig.~\ref{fig:ConfigurationSHZ} that the traffic jam involves regions of very high local density.  Since particle hops are almost impossible in such regions, swap moves play an important physical role in maintaining the ability of particles to move.
In models without swaps, there is a complete arrest of dynamics when the lattice is full; we view this as a lattice artefact, which is avoided by swap moves.  Nevertheless, we focus on small $\gamma=0.01,0.1$, so that particle motion is slow in very dense regions (even if they are not completely arrested).  It was shown in~\cite{korniss1997nonequilibrium} that significantly larger $\gamma$ destroys the perpendicular state and restores the homogeneous system. (If the presence of the other species does not slow particles down, then the traffic-jam mechanism does not operate.)

The traffic jam is a pronounced band of high-density oriented perpendicular to $\bm{E}$, so a suitable order parameter to measure these transitions is $\Phi(\bm{k}_\perp)$.  In fact, the charge-sensitive order parameter $\Psi(\bm{k}_\perp)$ has similar behaviour (data not shown), because the different particle species are separated within the high-density region.  We focus here on the $\Phi$ order parameter, for simplicity.

\subsection{\emph{Continuous} Transition}

We now discuss the transition from disordered to perpendicular states in more detail.  We separate the cases where the transition is continuous (larger $\rho$) and discontinuous (smaller $\rho$).

The  \emph{continuous} transition is illustrated in Fig.~\ref{fig:SORealOPSHZMain}.   In particular, Fig.~\ref{fig:SORealOPSHZMain}(a) shows the average of the order parameter $\langle \Phi(\bm{k}_\perp) \rangle$ as the field $\cal E$ is varied for several different system sizes.  For small fields, we find $ \langle \Phi(\bm{k}_\perp) \rangle=O(1/L)$ as expected in a homogeneous system. For larger fields,  $\langle \Phi(\bm{k}_\perp) \rangle$ increases continuously from this near-zero value, and its value is independent of $L$.  This is a signature of the perpendicular state shown in Fig.~\ref{fig:SORealOPSHZMain}.  To locate the position of the phase transition, we compute the scaled variance of the order parameter
\beq
S_\Phi(\bm{k}) = L_\parallel L_\perp \left[ \langle \Phi(\bm{k})^2 \rangle - \langle \Phi(\bm{k}) \rangle^2 \right],
\eeq
which is analogous to the structure factor of the fluid \cite{chaikin1995principles}.  Evaluating this quantity at the very small wavevector $\bm{k}_\perp$, gives a quantity analogous to the compressibility: it has a large peak at the transition, which is a signature of diverging fluctuations (critical behaviour).  
When identifying the phase boundary in Fig.~\ref{fig:PhaseDiagramSHZ}, we take the point where $S_\Phi(\bm{k}_\perp)$ is maximal, with an error bar corresponding to the separation of adjacent data points in Fig.~\ref{fig:SORealOPSHZMain}.

For an explicit demonstration of the spontaneous symmetry breaking, Fig.~\ref{fig:SORealOPSHZMain}(c) shows the probability distribution of the complex order parameter $P(\phi(\bm{k}_\perp))$.  In the disordered phase, the distribution of $\phi(\bm{k}_\perp)$ is a narrow peak centred at zero with a variance $O(1/L^2)$.  At the phase transition, this order parameter spreads out continuously onto a circle: the phase of this complex number reflects the position of the dense region within the system (and is random) but its modulus has a non-zero value.  This is the behaviour for the distribution of a two-component order parameter in a classical Landau theory of $\phi^4$ type, where the circle corresponds to the brim of a ``Mexican hat''.  Since the order parameter is complex, the symmetry breaking is of $U(1)$ type similar to \cite{dolezal2019large}.

Since our simulations are very long, we see that the system explores the 
full Goldstone mode in Fig.~\ref{fig:SORealOPSHZMain}(c), the position of the dense region fluctuates in time, and the phase of the complex ordered changes accordingly.  Still, the fact that $\Phi$ changes scaling from $O(1/L)$ to $O(1)$ shows that a dynamical phase transition is taking place.

\subsection{\emph{Discontinuous} Transition}

The \emph{discontinuous} transition for low $\rho$ is investigated in Fig.~\ref{fig:FORealOPSHZMain}. In particular, Fig.~\ref{fig:FORealOPSHZMain}(a) shows the average order parameter $\langle \Phi(\bm{k}_\perp) \rangle$, which clearly exhibits a jump discontinuity. 
At first-order (discontinuous) phase transitions one may expect hysteretic behaviour, where the behaviour of a system depends on its initial condition.  However, such finite lattice models are ergodic, so we expect that sufficiently long trajectories will eventually forget their initial conditions, allowing access to a well-defined steady state.  We checked that our simulations are long enough to achieve this by running simulations with both disordered and perpendicular initial conditions: our results for the order parameter are the same in both cases.  The order parameter shown in Fig.~\ref{fig:FORealOPSHZMain}(a) shows this steady state behaviour.  (The inset shows that almost identical results are obtained, independent of the initial condition, showing that the simulations are long enough to eliminate hysteresis.)

To explore this discontinuous transition, we consider relatively small systems and perform very long simulations (total time $\tau=5.8\times10^6$).  We adjust the field to $\cal E \approx$ $25$ for $\rho = 0.4$ so that the system is at coexistence between the homogeneous state and the perpendicular state.  
Fig.~\ref{fig:FORealOPSHZMain}(b) shows the probability distribution of the order parameter $\langle \Phi(\bm{k}_\perp) \rangle$ at phase boundary and Fig.~\ref{fig:FORealOPSHZMain}(c) shows a trajectory as the system switches between two metastable states.  The probability distribution of the order parameter has two peaks (corresponding to the two states), separated be a trough.  As the system size increases, the metastable states become increasingly well-defined and switches become increasingly rare.  This is accompanied by a deepening of the trough in the probability distribution of the order parameter.  For large system sizes, switches between the states become so rare  that a reliable estimation of this probability distribution is not possible; hence the data is only shown for $L\leq 35$.

The behaviour of the complex order parameter $\phi(\bm{k}_\perp)$ is illustrated in Fig.~\ref{fig:FORealOPSHZMain}(d).   As usual, the homogeneous phase is characterised by a narrow distribution close to zero.  Deep in the phase-separated state, the order parameter distribution forms a circle (the phase is random because the position of the dense region is random, but the modulus is well-defined and non-zero).  The characteristic signature of a first order transition (at $\cal E=$ 27.5) is that the order parameter distribution has two pieces: a sharp peak near zero (homogeneous state) as well as a circle (perpendicular state).  This corresponds to the behaviour of a two-component order parameter in a classical Landau theory of $\phi^6$ type \cite{goldenfeld2018lectures}.

\section{ELD model : parallel and perpendicular phase separation}
\label{sec:eld}

\subsection{ELD Model : motivation and definition}\label{sec:eld-def}

We have seen that the SHZ model supports phase separation into the perpendicular state of Fig.~\ref{fig:SchematicPT}(b).  
However, in off-lattice models of oppositely driven particles~\cite{dzubiella2002lane,klymko2016microscopic}, laning tends to occur more often, similar to the parallel state [Fig.~\ref{fig:SchematicPT}(c)].  
This difference indicates that lattice models fail to capture some aspects of the off-lattice systems.

It has even been suggested that the rigidity of the square lattice, combined with nearest neighbour hopping, is too restrictive for the laning effect (parallel state) to occur at all \cite{dzubiella2002lane, klymko2016microscopic}.  In the rest of this manuscript, we will introduce two square-lattice models where the parallel state does occur, the first of which is the ELD model.

\begin{figure}
    \centering
    \includegraphics[width=0.48\textwidth]{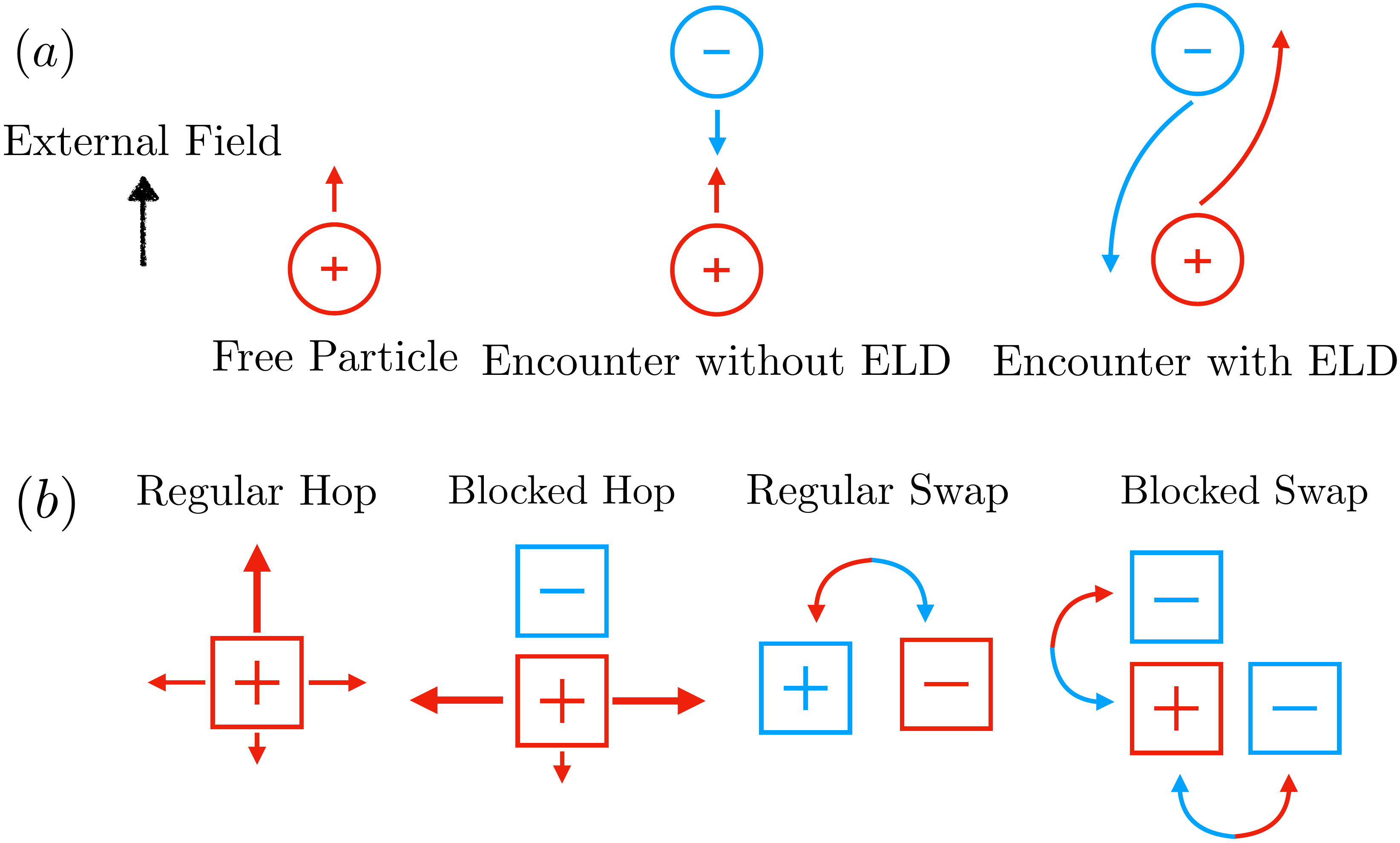}
    \caption{(a) Illustration of ELD mechanism, following~\cite{klymko2016microscopic}.  A particle is driven along the field direction and is diffusive in the lateral direction.  Without ELD, two opposite types of particles block each other when encountered.  With ELD, two opposite types of particles pass each other by diffusing about one diameter length in the lateral direction. (b) A schematic diagram of the dynamics of the ELD model. 
    A particle has different rates if it is blocked in its forward-moving direction by the opposite type of particle.
    }
    \label{fig:ELD} 
\end{figure}

To explain why the parallel state was not observed in previous models, we make two observations.
First, we emphasize a key observation of Klymko et al.~\cite{klymko2016microscopic}, that off-lattice particles experience enhanced lateral diffusion (ELD) if they collide with a particle of a different type, see Fig. \ref{fig:ELD}(a).  
The effect is reminiscent of some effects that occur in active matter -- the lateral motility of the particle depends strongly on its environment~\cite{cates2015motility}.  In the ELD model proposed here, this effect is included explicitly in the particle hopping rates, leading to an inherently non-equilibrium model that violates the local detailed balance formula (\ref{equ:SHZ-detbal}), similar to active systems~\cite{cates2013active, burger2016lane}.

The second observation is related to the Peclet number of a single driven particle in these models, defined as
\begin{equation}\label{eq:PecletForm}
    \text{Pe} = \frac{v_D \sigma}{D_0},
\end{equation}
where $v_D$ is the drift velocity of the particle, $\sigma=1$ the particle size, and $D_0$ the diffusion constant (measured at $\bm{E}=0$).  
For the SHZ model, we have $D_0=1$ and $v_D=1-{\rm e}^{-E}$ so that
\begin{equation}\label{eq:PecletSHZ}
    \text{Pe}_{\text{SHZ}} = 1-{\rm e}^{-E}.
\end{equation}
Hence, even if $E$ is very large, the maximal Pe is unity, see \cite{jack2008negative} for a similar issue in another lattice model. 
To observe laning effects in off-lattice models, one typically requires much larger Pe~\cite{dzubiella2002lane, klymko2016microscopic}. To encapsulate this, we design the ELD model such that the electric field enables an additional increase in the rate of forward hops. 

Given these considerations, we define the ELD model as follows.  The rate of motion of a particle depends on whether its driven direction along  $\bm{E}$ is ``blocked'' by an oppositely driven particle or not, see Fig.~\ref{fig:ELD}(b).  Particles that are not blocked have regular hops to adjacent empty sites with rate
\begin{equation}\label{eq:MIPSRegularHopRate}
W_{\text{ELD,RH}}(\bm{x},\bm{e},\sigma_{\bm{x}}) = \exp(\sigma_{\bm{x}}\bm{E}\cdot\bm{e}/2).
\end{equation}
These rates respect the local detailed balance relation (\ref{equ:SHZ-detbal}), but we note that if $\bm{e}$ is parallel to $\bm{E}$ then the forward rate may be very
large. This allows large Peclet numbers,
\begin{equation}\label{eq:PecletMIPS}
\text{Pe}_{\text{ELD}} = 2\sinh(E/2),
\end{equation}
which are not bounded by unity, unlike the SHZ case.

Similar to the SHZ model, we have regular swaps if a particle of the opposite type occupies a neighbouring site. This occurs with rate $W_{\text{ELD,RS}}(\bm{x},\bm{e},\sigma_{\bm{x}}) = \gamma W_{\text{ELD,RH}}(\bm{x},\bm{e},\sigma_{\bm{x}})$. This excludes neighbours along the driving direction, as those cases will fall in the blocked dynamics.  

\begin{table}
\begin{ruledtabular}
\begin{tabular}{lcccc}
 Move &RH&RS&BH&BS\\
\hline
Forward & $\exp(E/2)$& $\gamma \exp(E/2)$ & -- & $\gamma \exp(\alpha E/2)$ \\
Backward & $\exp(-E/2)$ & $\gamma \exp(-E/2)$ & $\exp(-E/2)$ & $\gamma \exp(-E/2)$ \\
Sideways  & $1$ & $\gamma$ & $\exp(E/2)$ & $\gamma \exp(E/2)$ \\
\end{tabular}
\end{ruledtabular}
\caption{\label{tab:ELDDynamicalRules}%
Dynamical transition rates of the ELD model (it is assumed here that $E>0$).  The forward move direction is in the $+y$ direction for red particles and the $-y$ direction for blue particles; backward and sideways moves are defined similarly.  The move types are regular hop (RH), regular swap (RS), blocked hop (BH) and blocked swap (BS),  see Fig.~\ref{fig:ELD}(b). Recall that ``blocked'' particles are those where the neighbouring site in the forward direction is occupied by a particle of the opposite species.
}
\end{table}

For blocked particles, the hop rates are
\begin{equation}\label{eq:MIPSSpecialHopRate}
W_{\text{ELD,BH}}(\bm{x},\bm{e},\sigma_{\bm{x}})  = \exp(g(\bm{e},E)/2),
\end{equation}
with 
\beq
g(\bm{e},E) =\begin{cases} 
      |E|, & \bm{e}  = \pm\hat{\bm{x}} \\
      -|E|, & \bm{e}  = -\sigma_{\bm{x}}\hat{\bm{E}} \\
      \alpha |E|, &\bm{e} = \sigma_{\bm{x}}\hat{\bm{E}} 
   \end{cases},
\eeq
where $\hat{\bm{E}}$ is a unit vector in direction of the electric field $\bm{E}$ and $\alpha$ is a parameter of the model, whose physical meaning will be discussed just below. 
The key point is that lateral hops ($\bm{e}  = \pm\hat{\bm{x}}$, perpendicular to $\bm{E}$) are strongly enhanced when particles are blocked.    
Similarly, we set
$W_{\text{ELD,BS}}(\bm{x},\bm{e},\sigma_{\bm{x}}) = \gamma W_{\text{ELD,BH}}(\bm{x},\bm{e},\sigma_{\bm{x}})$
 so that blocked swaps also experience ELD.  Detailed implementation of the ELD model is given in Sec~\ref{sec:app-eld-algo}.

Finally, we discuss the parameter $\alpha$.  Note that the case $\bm{e}=\sigma_{\bm{x}}\hat{\bm{y}}$ never appears for blocked hops: such transitions are always forbidden by the exclusion constraint.  This means that the parameter $\alpha$ is only relevant for swap moves: it governs the likelihood of particles swapping along the field, which promotes forward motion. Specifically, $\alpha$ sets the ratio of the driving field promoting lateral mobility (lower $\alpha$) or forwards motion (higher $\alpha$) in high-density regions. The dynamical rules of the ELD model are summarised in Table \ref{tab:ELDDynamicalRules}, see also Fig.~\ref{fig:ELD}(b).

\subsection{Overview and phase Diagram}

\begin{figure}
    \centering
    \includegraphics[width=0.48\textwidth]{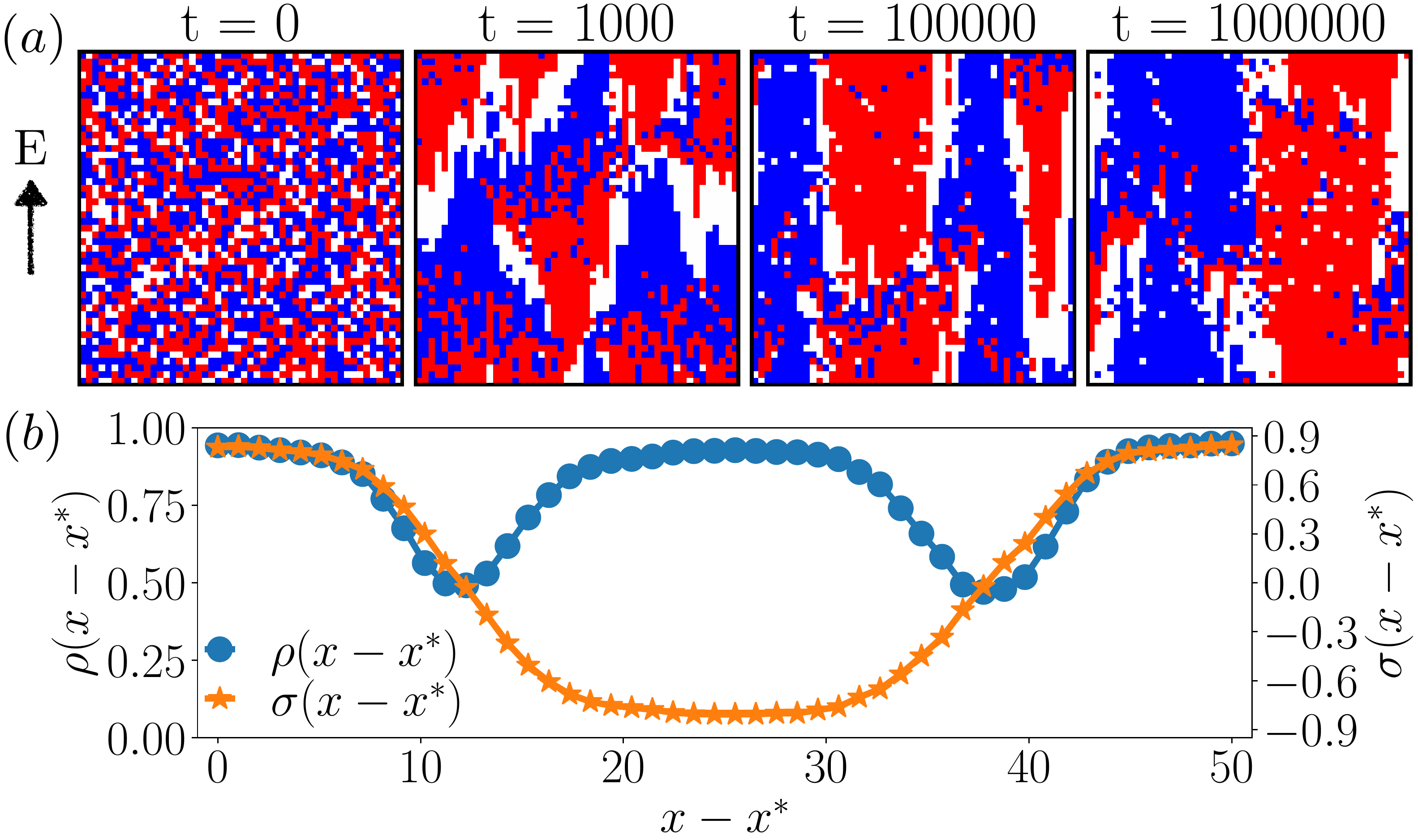}
    \caption{(a). Time series of configurations from a trajectory of the ELD model with $E=12$, $\rho=0.8$, $\gamma=0.1$, and $\alpha=0.7$ on a $50\times 50$ lattice. The ELD model exhibits phase separation parallel to the field.
    (b). Steady-state particle density $\rho(x-x^*)$ and charge density $\sigma(x-x^*)$ profile transverse to the driven direction.  The control parameters are the same as (a). 
    The density profiles are computed from a  single long trajectory, after the system reaches steady-state, the averaging time was $\tau = 4\times 10^5$. Charge density $\psi(\bm{k})$ is used to recenter the density profiles.
    }
    \label{fig:ConfigurationMIPS}
\end{figure}

\begin{figure}
    \centering
    \includegraphics[width=0.48\textwidth]{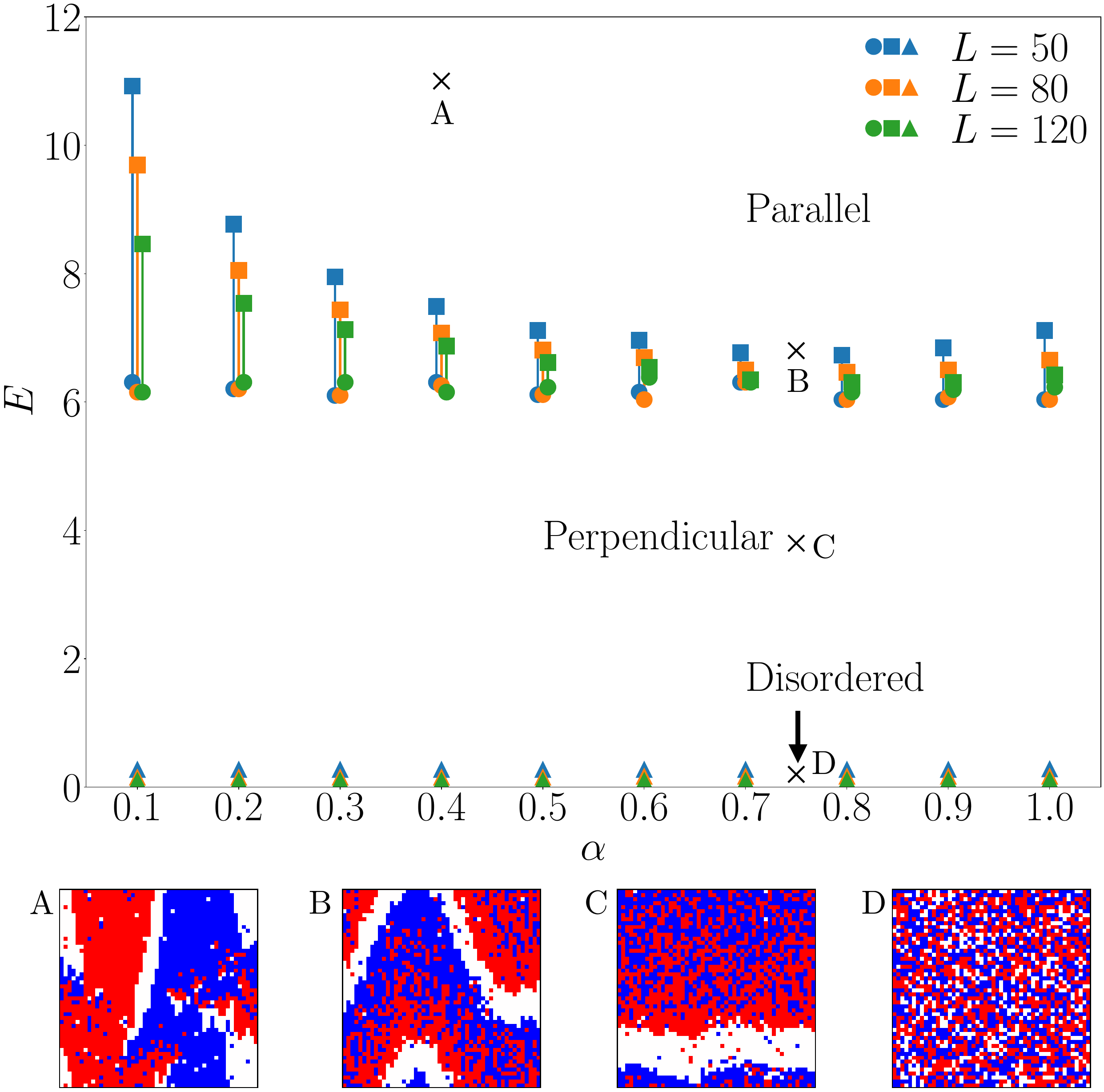}
    \caption{Phase diagram of the ELD model with 
    $\rho=0.8$, $\gamma=0.1$ for various system sizes.
    The phase diagram consists of disordered state, perpendicular state,  parallel state; in between the parallel and perpendicular is the ``zig-zag'' crossover regime.
    This regime is indicated by the vertical error bars (see the main text for a discussion).
    }
    \label{fig:PhaseDiagramMIPS}
\end{figure}

Fig.~\ref{fig:ConfigurationMIPS}(a) shows a time series of four configurations from the ELD model starting from a disordered initial condition: the system enters the parallel state that was sketched in Fig.~\ref{fig:SchematicPT}(c).  The phase separation can be seen from the steady-state density profiles of particles and charges in Fig.~\ref{fig:ConfigurationMIPS}(b), computed as in Sec.~\ref{sec:recen}.  The domains of red and blue particles are well-defined, but there are significant fluctuations when the domains are disrupted by particles of the minority phase. Fig.~\ref{fig:PhaseDiagramMIPS} shows the phase diagram for this model.  Note that the vertical axis is the field $E$ and not the scaled field ${\cal E}$, in contrast to the corresponding Fig.~\ref{fig:PhaseDiagramSHZ} for the SHZ model. In addition, parameter $\alpha$ is used as the horizontal axis, while we fix the density at the representative value $\rho=0.8$, similarly, $\gamma=0.1$. We run long simulations of $\tau=4\times 10^7$, and we discard the first half of the trajectory to ensure convergence to the steady state.  The methods used to identify the phase boundaries are described in Sec~\ref{sec:eld-dis-perp} and \ref{sec:eld-perp-para}.   

We observe that the ELD model exhibits both parallel and perpendicular states.  The transition from disordered state to perpendicular state occurs for $E=O(1/L)$ as in the SHZ model.  The parallel state is observed for much larger $E$, of order unity.  
(While we have not shown results for the SHZ model with very large $E$, that model only depends on $E$ through the factor ${\rm e}^{-E}$; once that quantity is close to zero, further increases in $E$ have little effect.) 
The ELD phase diagram shows a large region of uncertainty, between the parallel and perpendicular phases. 
This is due to strong finite-size effects that occur close to the transition, associated with different kinds of phase separation. 
The resulting ``zig-zag'' states are discussed in Sec.~\ref{sec:eld-perp-para} below.

Before discussing the phase transitions in detail, we briefly discuss the dependence of the system on other model parameters.  We restrict $\alpha\leq1$.  (Larger $\alpha$ corresponds to an unphysical regime where forward motion is enhanced when particles are blocked.)  Also, the ELD mechanism relies on collisions between particles, so it operates most effectively at high density, and the parallel state is suppressed in states with much smaller $\rho$ -- we consider $\rho=0.8$ which is representative of the regime where the parallel state can be found.  Lastly, we recall that the parameter $\gamma$ determines the relative rate of swap moves. It is desirable that $\gamma$ is small to enforce the physical idea that particle mobility is slow in regions of high density.  However, very small values of $\gamma$ can lead to slow dynamics and inefficient simulations.  Hence, we use $\gamma=0.1$.

\subsection{Transition between disordered and perpendicular state}\label{sec:eld-dis-perp}

\begin{figure}
    \centering
    \includegraphics[width=0.44\textwidth]{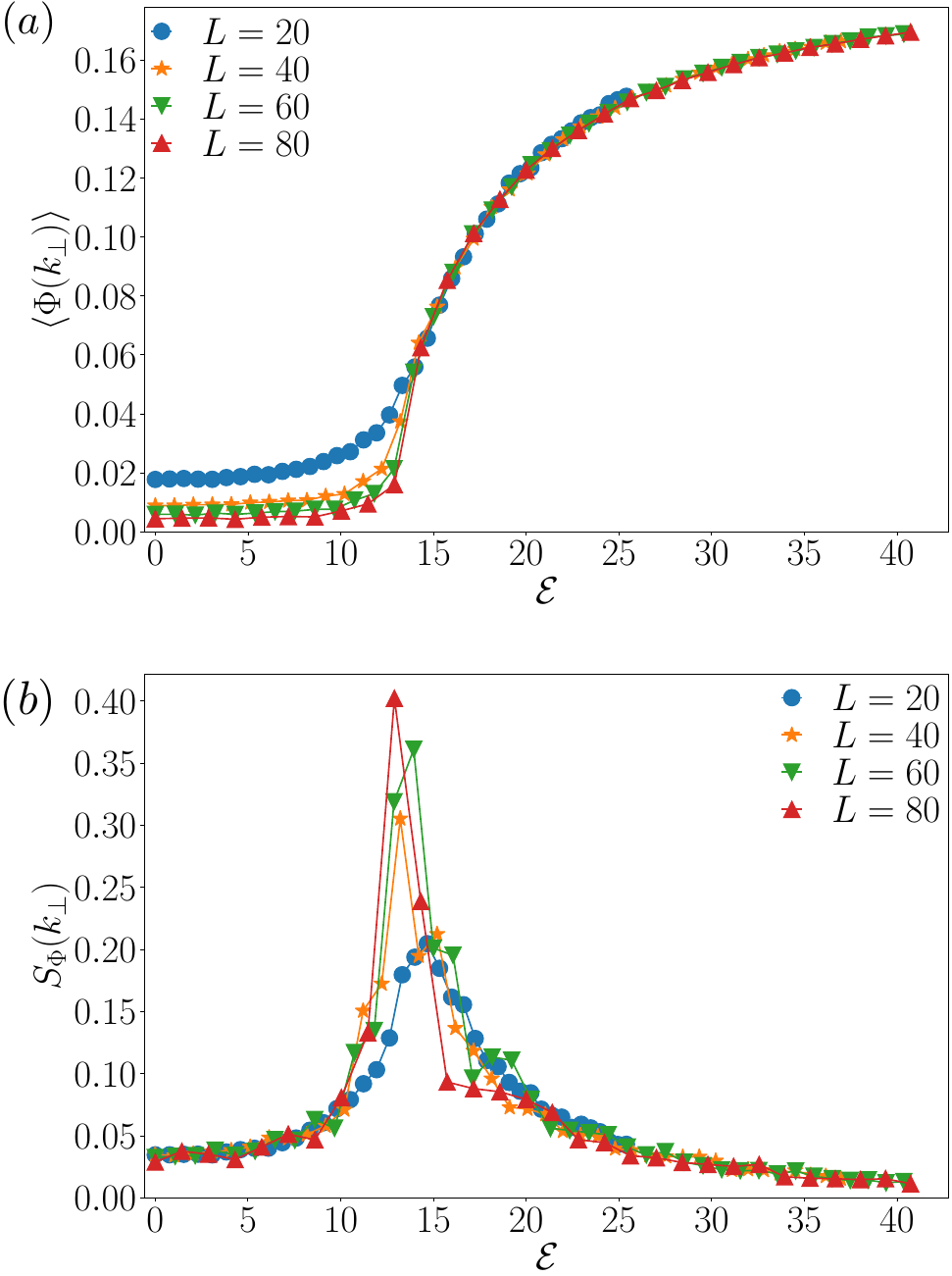}
    \caption{(a). Order parameter $\Phi(\bm{k}_\perp)$ vs $\cal E$. (b). Scaled variance of the order parameter $S_{\Phi}(\bm{k}_\perp$) vs $\cal E$ for the disordered-perpendicular transition for four different system sizes in the ELD model with $\rho=0.8$, $\gamma = 0.1$, and $\alpha=0.7$.
    }%
    \label{fig:SORealOPMIPSMain}%
\end{figure}

For small $E$, the ELD model is similar to the SHZ model.  One observes a similar transition from the homogeneous state to the perpendicular state for ${\cal E}=O(1)$, corresponding to $E=O(1/L)$.
Fig.~\ref{fig:SORealOPMIPSMain} illustrates this behaviour; it is a continuous transition, similar to Figs.~\ref{fig:SORealOPSHZMain}(a,b), and the data collapses as a function of the same $\cal E$ parameter that was used in the SHZ case.  The corresponding phase boundary in Fig.~\ref{fig:PhaseDiagramMIPS} was identified via the peak in $S_\Phi$ similar to the SHZ model.
We also confirmed that for lower densities, the disorder-perpendicular transition becomes discontinuous (data not shown). 

\subsection{Transition between parallel and perpendicular states}\label{sec:eld-perp-para}

\begin{figure}
    \centering
    \includegraphics[width=0.44\textwidth]{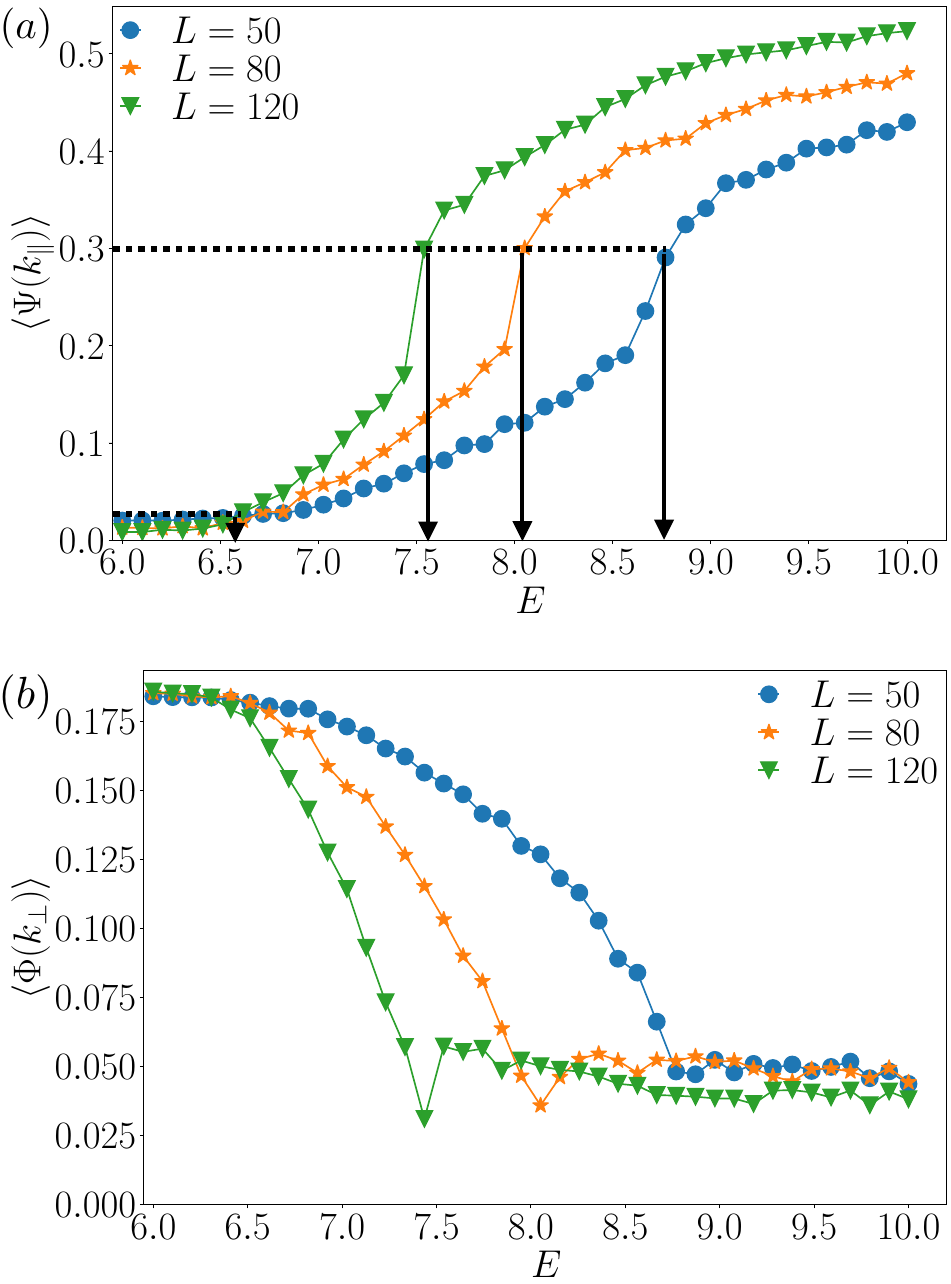}
    \caption{(a). Order parameter  $\Psi(\bm{k}_{\parallel})$ vs $E$.  (b). Order parameter $\Phi(\bm{k}_\perp)$ vs $E$ for the perpendicular-parallel transition for four different system sizes in the ELD model with $\rho=0.8$, $\gamma = 0.1$, and $\alpha = 0.2$. Hysteresis is not observed at this transition.  The dotted lines in (a) indicate the minimum/maximum order parameter values where the system is still in the ``zig-zag'' state.
    }%
    \label{fig:FORealOPMIPSMain}%
\end{figure}

Next, we consider the behaviour of the ELD model at larger $E$, and the transition between perpendicular and parallel states.  In Fig.~\ref{fig:PhaseDiagramMIPS}, we see that a direct transition from disordered to parallel state is not observed, as there is always an intermediate perpendicular state. To study the parallel state, we use the order parameter $\Psi(\bm{k}_\parallel)$. This order parameter focuses on ordering of the charges which separate in the parallel state. (The density order parameter $\Phi(\bm{k}_\parallel)$ does not provide a clear signal of this transition).  On the other hand, $\Phi(\bm{k}_\perp)$ is the most suitable order parameter for the perpendicular state, in which the charge-ordering is weaker (at least for $\gamma=0.1$, as considered here).

Fig.~\ref{fig:FORealOPMIPSMain} shows the order parameters for parallel and perpendicular phases across the transition.
The parallel order parameter $\Psi(\bm{k}_\parallel)$ shows a jump, after which it gradually increases.  Similar behaviour was observed in experiments on laning~\cite{vissers2011band} with time-dependent electric fields.  At the same time, the perpendicular order parameter $\Phi(\bm{k}_\perp)$ decreases.

\begin{figure}
    \centering
    \includegraphics[width=0.48\textwidth]{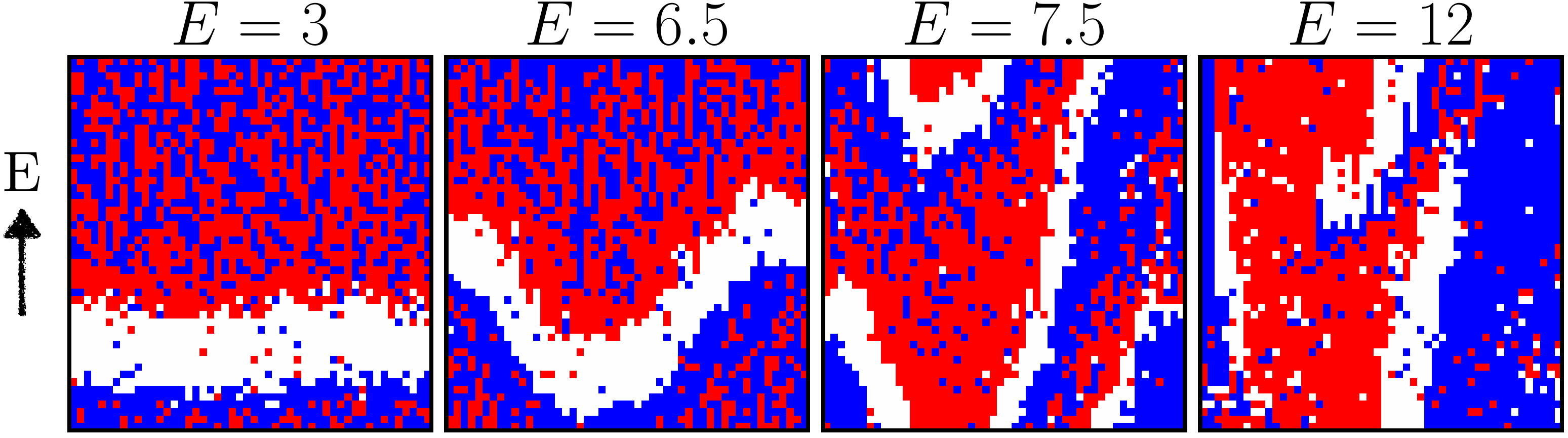}
    \caption{Steady state behaviour of the ELD model with different electric field strengths at $\rho = 0.8$, $\gamma = 0.1$, and $\alpha = 0.7$ on a $50\times 50$ lattice.  As the electric field increases, the particle-hole interface
    bends and the system enters a  crossover regime between the perpendicular-parallel transition. We call this steady state the ``zig-zag" state.}
    \label{fig:ZigZag}
\end{figure}

Here we briefly explain how we obtained the vertical errorbars' bounds in Fig.~\ref{fig:PhaseDiagramMIPS} for the perpendicular-parallel transition. 
We identify the order parameter values at which the order parameter of our largest simulation box makes an abrupt jump [the dotted lines shown in Fig.~\ref{fig:FORealOPMIPSMain}(a)]. 
We then use these order parameters values to identify the corresponding electric field value, which bounds this transition regime for smaller system sizes.   We say that we have well-defined perpendicular and parallel phases outside these two dotted lines. In between these lines, we find a transition regime, which we call a ``zig-zag'' state (Fig.~\ref{fig:ZigZag})

In this regime, the perpendicular state no longer consists of a straight band across the system.  Instead it bends, forming a characteristic zig-zag shape that becomes increasingly pronounced as $E$ increases.  Eventually, the shape becomes unstable and breaks into a parallel state.  Note that some states with zig-zag bands were also observed in experiments~\cite{leunissen2005ionic, vissers2011band}, but it is not clear if these represent non-equilibrium steady states or metastable (transient) behaviour.

We make two observations about this complex transition.  First, Figs.~\ref{fig:PhaseDiagramMIPS} and~\ref{fig:FORealOPMIPSMain} both show that the zig-zag region gets narrower with increasing system size.  We suspect that the zig-zag state is stabilised by finite-size effects so that only parallel and perpendicular states will survive as $L\to\infty$, although further work would be required to confirm this beyond doubt.  Second, in contrast to the discontinuous transition of the SHZ model (Fig.~\ref{fig:FORealOPSHZMain}), we do not observe significant hysteresis in the parallel-perpendicular transition of the ELD model, despite the jump in the order parameter.  On increasing $E$, the observed behaviour is more consistent with a linear instability of the perpendicular state, i.e. forming the zig-zag, followed by a second instability of the zig-zag, i.e. forming the parallel state.  More specifically, starting from the perpendicular state with broken symmetry along the $y$ direction, the first instability corresponds to breaking of translational symmetry along the $x$-direction (zig-zag), and the second to restoration of translational symmetry along the $y$-direction (parallel state).

As a more general point, this transition is not a spontaneous symmetry breaking but rather a transition between two symmetry-broken states.  As such, it is not surprising that it does not fit naturally into a classical Landau theory, in contrast to the behaviour of the SHZ model in Figs.~\ref{fig:SORealOPSHZMain} and \ref{fig:FORealOPSHZMain}, and should be understood from a linear stability perspective.

\subsection{Linear Stability Analysis}\label{sec:eld-linear}
We briefly describe a schematic analysis of the instabilities of the homogeneous state in this system.  
Based on~\cite{korniss1997nonequilibrium}, consider a hydrodynamic theory where $\rho_\sigma(\bm{x})$ is the local density of particles with charge $\sigma=\pm$.  Also let $\rho_{\rm T}=\rho_+ + \rho_-$ be the total density.  This is normalised such that $\rho_{\rm T}=1$ corresponds to a completely filled lattice.  Then a simple theory for the SHZ model is 
\beq \label{eq:HydrodynamicEq}
\frac{\partial \rho_\sigma}{\partial t} = - \sigma \epsilon \nabla_y [ \rho_\sigma (1-\rho_{\rm T}) ] + \nabla^2 ( D\rho_\sigma),
\eeq
with $\epsilon = 2 \tanh(E/2)$ and $D$ a diffusion constant. 
This equation can be derived using a mean-field approximation~\cite{korniss1997nonequilibrium}. It is possible to derive more accurate hydrodynamic descriptions~\cite{mason2022macroscopic} but the mean-field approximation is sufficient to capture the essential physics.

Now consider a perturbation about an inhomogeneous state 
\begin{equation}\label{eq:Perturbation}
    \rho_\sigma(\bm{x},t) = (\rho/2) + A_\sigma (\bm{k}) e^{-\lambda t+i\bm{k}\cdot \bm{x}}.
\end{equation}
where $\bm{k}$ is the wavevector of the perturbation, and $\lambda$ its decay rate.  Substituting (\ref{eq:Perturbation}) into (\ref{eq:HydrodynamicEq}), one finds an instability (that is, $\lambda<0$) for sufficiently large fields, at wavevector $\bm{k}_\perp$.  Specifically, the system is unstable if $2\rho>1$ and
\beq
   \epsilon > \epsilon^* =\frac{2\pi D}{L_{\parallel}}\sqrt{\frac{1}{(1-\rho)(2\rho-1)}}.
   \label{equ:eps-star}
\eeq
For $2\rho<1$, this simple theory predicts that homogeneous state is always stable, 
The above linear stability calculation is consistent with the qualitative results that the particles in the SHZ model develop a macroscopic band perpendicular to the external field.
Quantitative agreement between theory and simulation is not expected, because of the simplicity of the theoretical description.  In fact, it remains a challenging problem to derive quantitatively accurate descriptions of the hydrodynamic behaviour of such models~\cite{mason2022macroscopic}.
 
For a minimal description of ELD, note that diffusion of $+$ particles in the $x$ direction is much larger in regions where the density of the $-$ particles is large.  This effect may be captured by an anisotropic theory similar to MIPS \cite{cates2015motility}. For example:
\beq \label{eq:HydrodynamicEqMIPS}
\begin{gathered}
     \frac{\partial \rho_+}{\partial t} = -\epsilon\partial_y[\rho_+(1-\rho_{\rm T})] + \nabla_y^2 [  \rho_+ D_{yy} ]       
     +\nabla_x^2 [ \rho_+ D_{xx}(\rho_-) ] 
     \\
     \frac{\partial \rho_-}{\partial t} = +\epsilon\partial_y[\rho_-(1-\rho_{\rm T})] + \nabla_y^2 [  \rho_- D_{yy} ]       
     +\nabla_x^2 [ \rho_- D_{xx}(\rho_+) ] 
     \end{gathered}
\eeq
where $D_{yy}$ is a constant diffusivity in the $y$-direction, while the diffusivity $D_{xx}$ in the $x$-direction depends on the concentration of the other species, see~\cite{chakrabarti2004reentrance, frusawa2022stochastic} for a similar idea.  For the ELD model, it is natural that $\epsilon = 2\tanh(E/2)$, similar to (\ref{eq:PecletMIPS}). 
The enhancement of lateral diffusion is exponential in $E$, and this effect is pronounced when $\rho_\sigma$ is large, so 
a crude estimate for $D_{xx}$ might be $ D_{xx}(\rho_\sigma) \approx D(1 + \rho^{2}_{\sigma} \cosh(E/2))$.  

Homogeneous states in this model still show an instability to the perpendicular state, as in SHZ.  However, there is also a MIPS-type instability to a parallel state: using again (\ref{eq:Perturbation}), it takes place when
\begin{equation}
    \frac{2}{\rho}<\frac{|D^{\prime}_{xx}(\rho/2)|}{D_{xx}(\rho/2)}.
    \label{equ:mips-instability}
\end{equation}
The above linear stability analysis (details given in Appendix \ref{sec:Stability_analysis}) suggests that the instabilities do not depend on the wavevector. Instabilities occur at all scales, and the instabilities with large wavevectors grow faster than the ones corresponding to small wavevectors. 
Such a feature is consistent with the fact that particles first form multiple traffic lanes parallel to the external field in a sufficiently large system. Traffic lanes then coarsen into larger domains and eventually phase separate parallel to external field.

In practice, it is clear from the phase diagram that the first instability (on increasing $E$ from zero) is to the perpendicular state. This is also consistent with the linear stability analysis. Eq. (\ref{equ:eps-star}) predicts that the transition to the perpendicular state occurs for $E=O(1/L)$. The MIPS instability requires $E=O(1)$. 
When this instability occurs, the MIPS mechanism will operate in the dense region of the perpendicular state. Almost all particles are blocked in that region, so lateral diffusion will be strongly enhanced in that case.
Indeed, the instability to the zig-zag state shares many features with a traditional linear MIPS instability (wavevector $\bm{k}_\parallel$ and the formation of red-rich and blue-rich regions at the upward and downward-pointing ``tips'' of the zig-zag).

\subsection{Discussion :  ELD model}
\label{sec:eld-discuss}

We summarise the results presented so far. 
The behaviour of the ELD model can be rationalised by considering two effects.  For sufficiently large density, the perpendicular phase-separated state appears at very small fields $E=O(1/L)$, similar to the SHZ model discussed in \ref{sec:shz-phase}. Recalling~\eqref{equ:eps-star}, this transition can be rationalised as a linear instability of the hydrodynamic equations: if the density increases locally, then particles tend to be blocked. This reduces the effectiveness of the field $E$ in driving a current.  Hence, particles slow down in the high-density region, while particles arrive quickly from the low-density region.
This leads to a ``traffic jam'', stabilising perpendicular states.  In off-lattice models, similar effects occur in ``freezing by heating'' models~\cite{helbing2000freezing}.

On the other hand, the parallel phase-separated state appears in the ELD model but not in the SHZ.  For the ELD model, blocked particles move quickly in the lateral direction: blocking only operates between particles of different species. This drives particles into regions where their own species predominate.  This effect is analogous to MIPS~\cite{cates2015motility}, as indicated by the (qualitative) stability analysis that leads to \eqref{equ:mips-instability}.  It leads to the parallel phase-separated state, which also appears in (off-lattice) models of laning~\cite{dzubiella2002lane, lowen2003nonequilibrium, chakrabarti2003dynamical, chakrabarti2004reentrance, glanz2012nature, kohl2012microscopic, klymko2016microscopic, wachtler2016lane, oliveira2016keep, poncet2017universal,reichhardt2018velocity,reichhardt2018laning,geigenfeind2020superadiabatic}.
Moreover, the ELD effect is strongest in high-density regions, where the blocking is most pronounced.  
This leads to an interplay between  parallel and perpendicular phase separation, and the instability to parallel phase separation emerges inside a high-density region that has already occurred via perpendicular phase separation. 
It seems likely that this interplay is the origin of the zig-zag state.

\section{Phase Transition in TKLS model}
\label{sec:tkls}

\subsection{Motivation and model definition}

\begin{figure}
    \centering
    \includegraphics[width=0.48\textwidth]{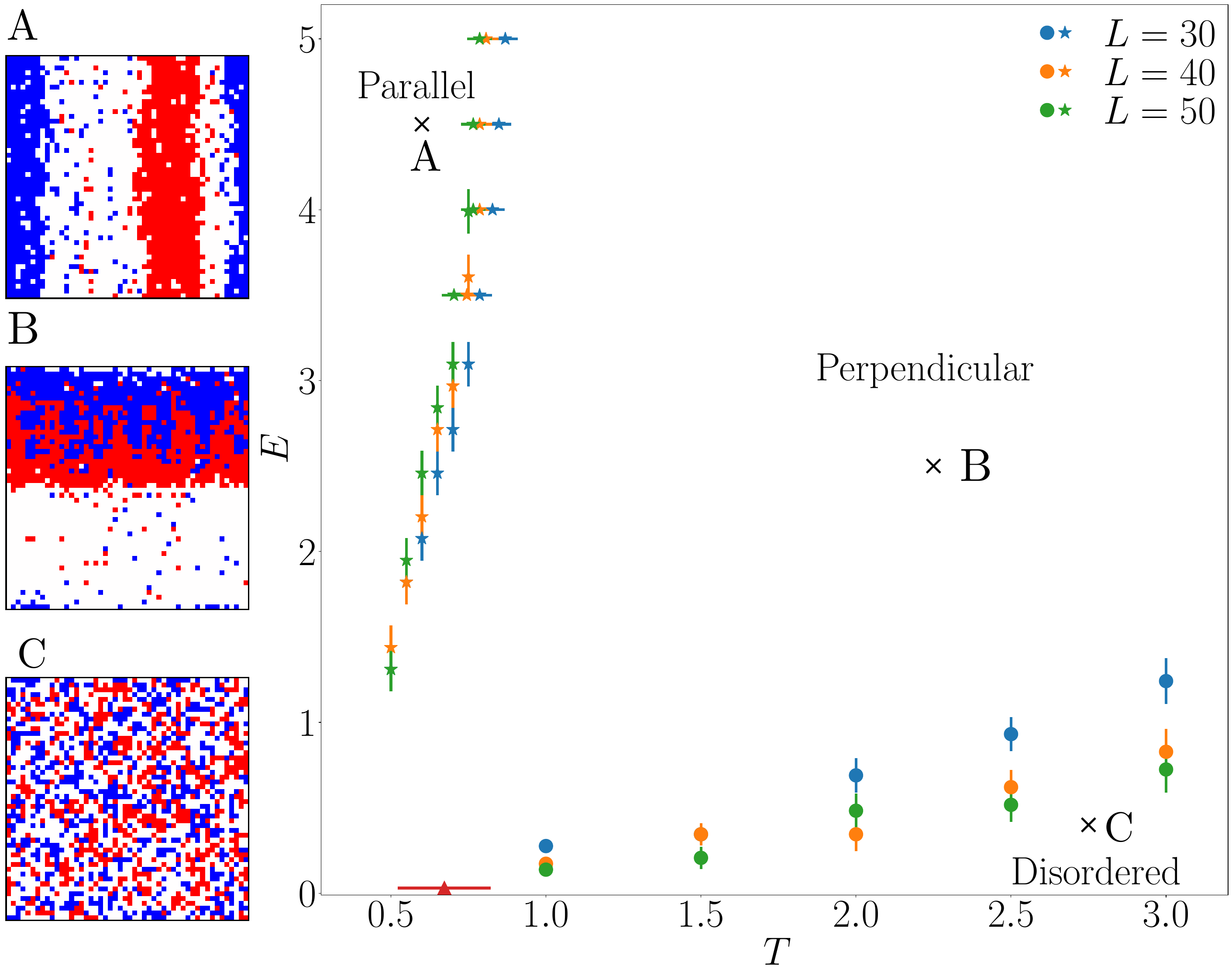}
    \caption{Phase diagram of the TKLS model for various system sizes with $\rho=0.5$, $\gamma=0.05$. Three states are observed: disordered state, perpendicular state, parallel state. The equilibrium transition ($E=0$) occurs around $T_c\approx0.7\pm0.2$.}
    \label{fig:PhaseDiagramTKLS}
\end{figure}

The enhanced lateral diffusion mechanism suggests that the diffusion constant of a particle is large when it is surrounded by oppositely moving particles and is small when surrounded by the same type of particles. Such environment-dependent diffusion can be thought of as an effective repulsion between particles of opposite types. We analyse a model where similar interactions are introduced directly (a variation of the KLS model) to verify this observation. Specifically, we introduce repulsive interactions between opposite types, and also an attraction between particles of the same type.

The resulting model also supports both parallel and perpendicular states. This indicates that a combination of effective interactions and external driving is the origin of the behaviour observed in the ELD model, consistent with earlier ideas about mechanisms for laning effects~\cite{klymko2016microscopic,PhysRevE.74.011403}.

The original KLS model has a single particle species with attractive interactions and external driving~\cite{katz1984nonequilibrium}. The model introduced here is a two-species variant of this model, hence we call it the two-species KLS (or TKLS) model.
The interaction energy of the TKLS model is
\begin{equation}\label{eq:NNI}
H = -J\sum_{\langle\bm{x},\bm{x}^{'}\rangle}\sigma_{\bm{x}}\sigma_{\bm{x}^{'}},
\end{equation}
where $J$ is the interaction strength and the sum runs over distinct pairs of neighbouring lattice sites.
There is no distinction between blocked and unblocked particles: 
particles hop with rate  
\begin{equation}\label{eq:TKLSphRate}
W_{\text{H,TKLS}}(\bm{x},\bm{e},\sigma_{\bm{x}}) = \min(1, \exp[(\sigma_{\boldsymbol{x}}\bm{E}\cdot\bm{e}-\Delta H )/T]),
\end{equation}
where $\Delta H$ is the change in interaction energy $H$ associated with the hop and $T$ is the temperature.  Similarly, we take the swap rate as
\begin{equation}\label{eq:TKLSppRate}
W_{\text{S,TKLS}}(\bm{x},\bm{e},\sigma_{\bm{x}}) = \gamma\min(1, \exp[(2\sigma_{\boldsymbol{x}}\bm{E}\cdot\bm{e}-\Delta H)/T]) \; .
\end{equation}

One recovers the SHZ model by setting $T=1$ and $J=0$.  In the following, we set $J=1$. This does not lose any generality as the dependence on the three parameters $(T,J,E)$ is fully determined by the two ratios $T/J$ and $E/T$. Note, however, that the parameter $E$ of the SHZ and ELD models corresponds to $E/T$ in the TKLS model, so it is useful to define the analogue of ${\cal E}$ in this model as
\beq
{\cal E}_{\rm TKLS} = 2 L_{\parallel} \tanh{(E/2T)} \;  .
\eeq
A similar model was studied in~\cite{lyman2002phase, lyman2005steady}, but all interactions were attractive in that case (and there were no swap moves). In this case, the parallel phase was not observed.  Connections between the behaviour of KLS and TKLS models are discussed in Sec.~\ref{sec:discuss-tkls}.

\subsection{Phenomenon and Phase diagram}\label{sec:tkls-phase}

\begin{figure}
    \centering
    \includegraphics[width=0.48\textwidth]{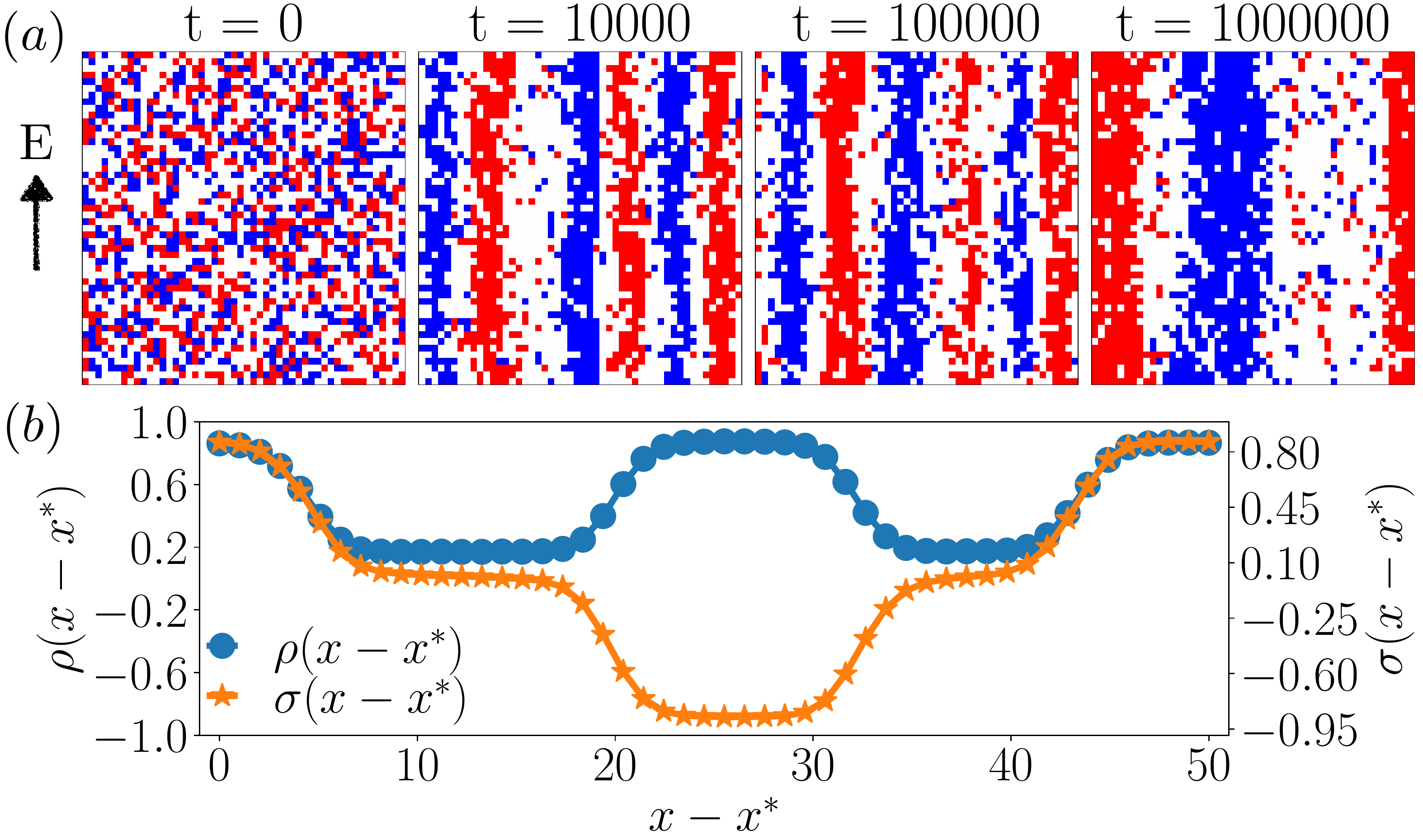}
    \caption{(a). Time series of configurations from a trajectory of the TKLS model with $E=5$, $\rho=0.5$, $\gamma=0.05$, and $T=0.7$ on a $50\times50$ lattice. The system exhibits a phase transition between the disordered phase and the parallel phase. After lanes are formed, a coarsening over time is observed.
    (b)~Steady-state particle density $\rho(x-x^*)$ and charge density $\sigma(x-x^*)$ profile transverse to the driven direction with same control parameters as in (a). The density profiles are obtained by averaging over $\tau = 4\times 10^5$ for a single long trajectory, after the system has reached its steady-state. Charge density $\psi(\bm{k})$ is used to recenter the density profiles.
    }
    \label{fig:ConfigurationTKLS}
\end{figure}

Fig.~\ref{fig:PhaseDiagramTKLS} shows the TKLS phase diagram in the $(T,E)$ plane, for representative parameters $\gamma=0.05$ and $\rho=0.5$.  For $E=0$,  the general behaviour of this system is known from the Blume-Capel model \cite{zierenberg2017scaling, butera2018blume} -- the system is homogeneous at high temperature, it phase separates at low temperature into a state of three-phase coexistence (two phases that consist of predominately red and blue phases, coexisting with a dilute vapour).  For $\rho=0.5$ (as discussed here), this transition takes place by crossing a binodal whose position we estimate as $T\approx0.7\pm0.2$. 

As in other sections, we focus on non-equilibrium steady states with $E>0$.  
In this region, the phase behaviour must be deduced from long dynamical simulations.  
For low temperatures and small $E$, simulations indicate a complicated interplay between equilibrium and non-equilibrium transitions.  At these densities, the equilibrium state of phase coexistence usually has the particles in circular droplets that do not cross the periodic boundaries, but transitions into morphologies similar to Fig.~\ref{fig:SchematicPT} can already occur at equilibrium \cite{mayer1965interfacial, leung1990geometrically}.  To avoid the slow dynamics associated with transitions between these morphologies (and noting that sampling can also be frustrated by low acceptance rates when $T$ is small), we focus here on relatively high temperatures. 
For the values of $E$ that we do consider, the phase diagram shows both parallel and perpendicular states, including transitions from the disordered state to perpendicular state, and from the perpendicular state to the parallel state. 
This general phenomenology is similar to the ELD model and we remark on the similarities when discussing the transitions. 

\begin{figure}
    \centering
    \includegraphics[width=0.44\textwidth]{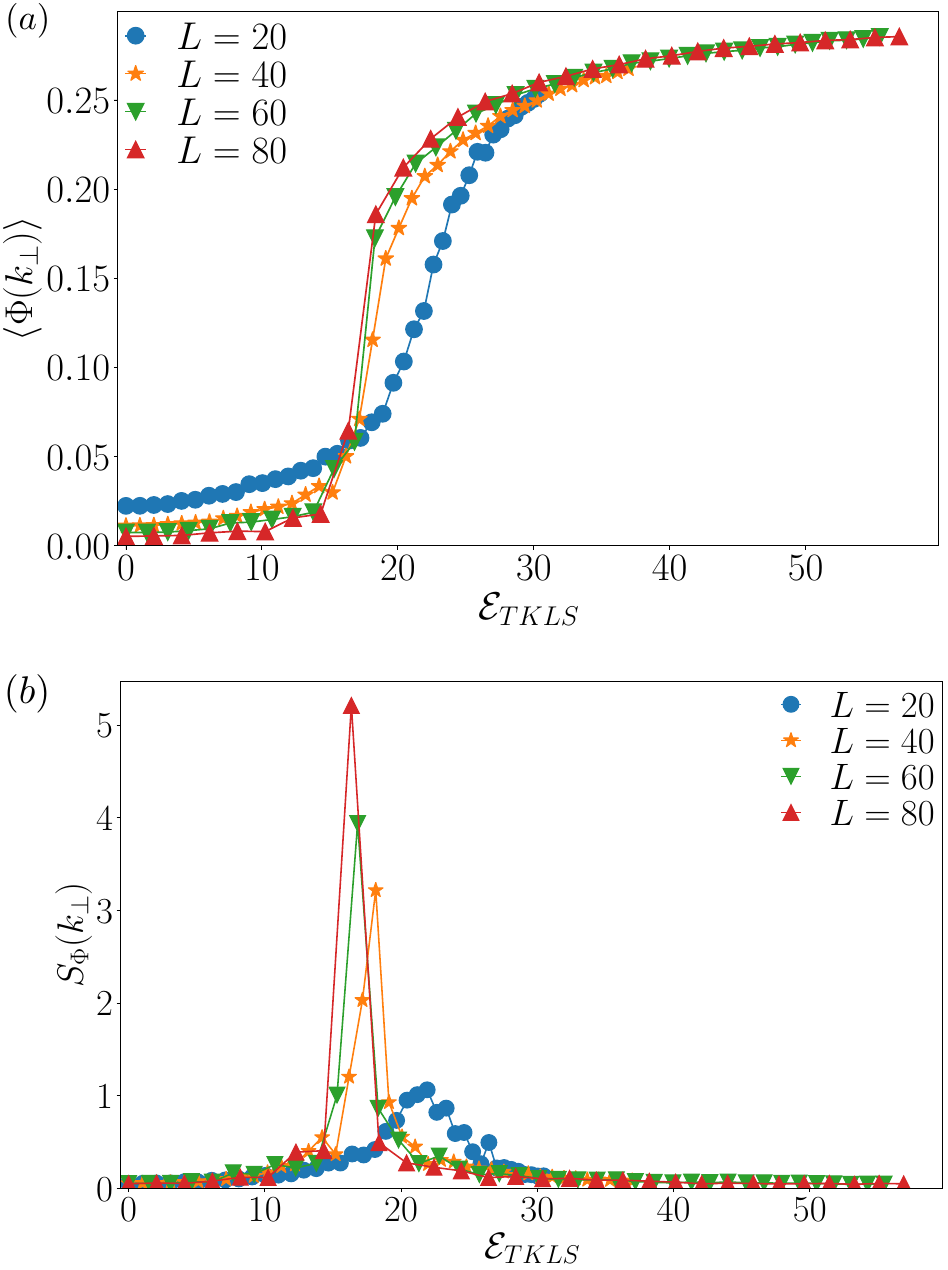}
    \caption{
    (a)~Order parameter $\Phi(\bm{k}_\perp)$ vs $\cal E_{\text{TKLS}}$. (b)~Scaled variance of the order parameter $S_{\Phi}(\bm{k}_\perp$) vs $\mathcal{E}_{\text{TKLS}}$ for the disordered-perpendicular for four different system sizes with $\rho=0.5$, $\gamma = 0.05$, and $T=5$ in the TKLS model.
    }% 
    \label{fig:SORealOPTKLSMain}%
\end{figure}

\begin{figure*}
    \centering
    \includegraphics[width=0.87\textwidth]{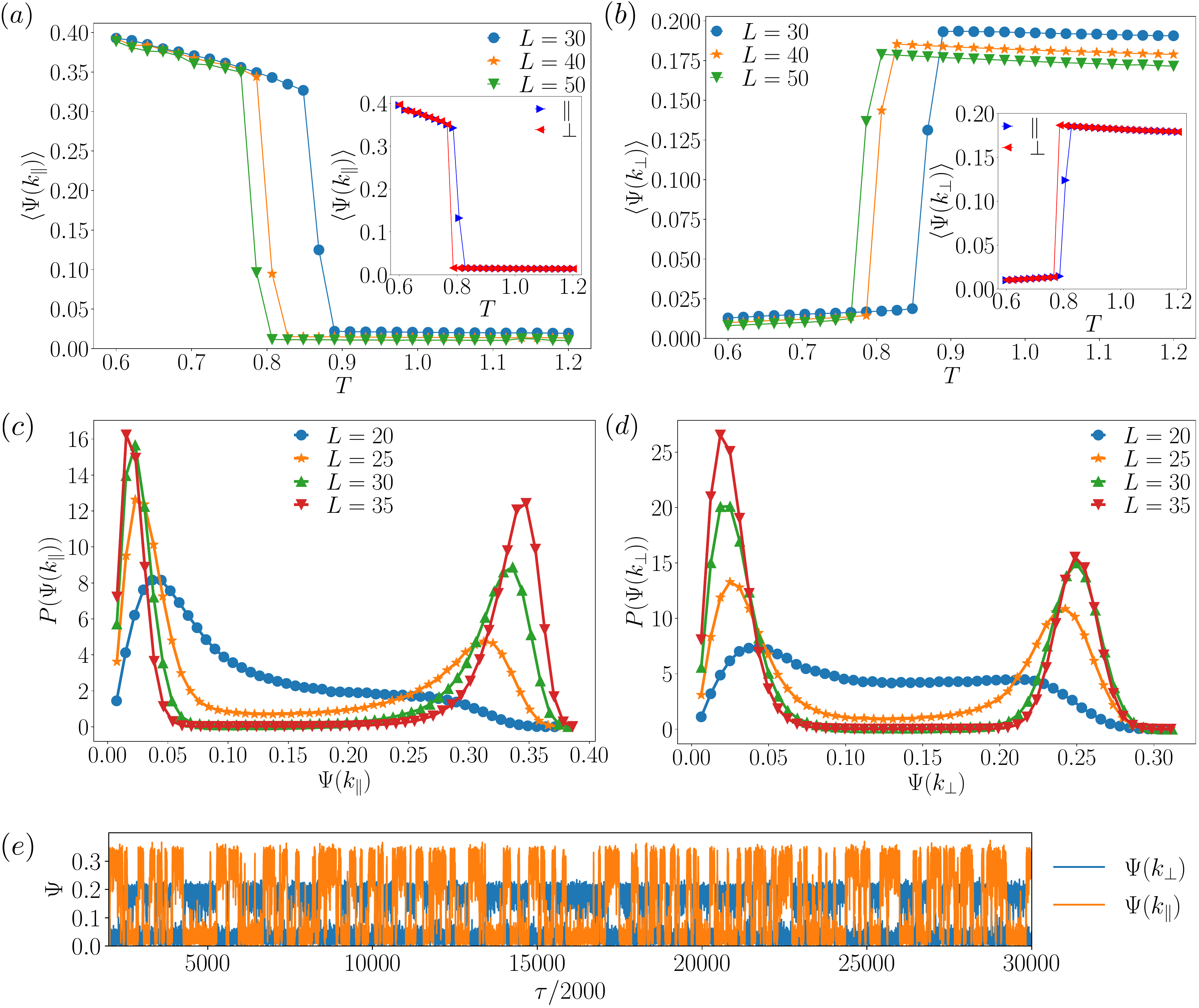}
    \caption{
    (a)~$\Psi(\bm{k}_{\parallel})$ vs $T$.  (b)~$\Psi(\bm{k}_\perp)$ vs $T$ for three different system sizes for $E = 5$, $\rho = 0.5$, and $\gamma=0.05$ for the perpendicular-parallel transition. The insets show that these results do not depend on the initial condition: results are shown for simulations starting in parallel ($\parallel$) or perpendicular states ($\perp$), which show almost identical behavior.  Each point is an independent simulation. Data are shown for $L=40$.
    (c,d)~Probability distributions of the order parameters shown in panels (a,b) at four different system sizes. (e)~Time series of the order parameters shown in panels (a,b) for system size $L_{\parallel}=L_{\perp}=25$.
    }%
    \label{fig:FORealOPTKLSMain}%
\end{figure*}

The following subsections will analyse the transitions.  As a preliminary for that analysis, Fig.~\ref{fig:ConfigurationTKLS}(a) shows how the system evolves from a  disordered initial condition into the parallel state.  The steady-state profiles for the parallel state in Fig.~\ref{fig:ConfigurationTKLS}(b) resemble the profiles of the ELD model [Fig.~\ref{fig:ConfigurationMIPS}(b)] although the interfaces between the phases are better defined, and there are fewer fluctuations of the minority particles within the phases. The coarsening processes are also different in the two models. In the TKLS model, a multi-domain parallel state forms initially, which then coarsens to full phase separation.  This is quite different from the analogous process in the ELD model (Fig.~\ref{fig:ConfigurationMIPS}), where the system initially forms a state of perpendicular phase separation, which becomes unstable and forms the parallel phase.  Indeed, the following analysis will show that while the steady states in the ELD and TKLS models are similar, the transition between parallel and perpendicular states is quite different.

\subsection{Transition between disordered and perpendicular states}\label{sec:tkls-dis-perp}
We first analyse the transition from a disordered state to the perpendicular state, upon increasing $E$ from zero.  We note the disordered-perpendicular transition only happens with high temperature, i.e. $T\gtrapprox1$. 
Besides that, the transition from disordered to perpendicular is very similar to the one that occurs in ELD and SHZ models, as it takes places at $E=O(1/L)$. 
[See Fig.~\ref{fig:SORealOPTKLSMain}, which is analogous to Fig.~\ref{fig:SORealOPMIPSMain} and Fig.~\ref{fig:SORealOPSHZMain}(a,b).]

\subsection{Transition between parallel and perpendicular states}\label{sec:tkls-perp-para}

The transition between perpendicular and parallel states in the TKLS model is quite different from the transition in the ELD case.
Results are shown in Fig.~\ref{fig:FORealOPTKLSMain}: the transition is studied by reducing $T$ at fixed field $E$. Fig.~\ref{fig:FORealOPTKLSMain}(a,b) shows that the order parameters experience a discontinuous jump at the transition. The insets show that almost identical results are obtained, independent of the initial condition, showing that the simulations are long enough to eliminate hysteresis.
We obtained the histograms by running the simulations up to $\tau=12\times 10^{10}$ and discard the first $4\times 10^7$. This ensures the system switches multiple times between two states and the area under two peaks converges. 

From the behaviour of the order parameters and their probability distribution, we can conclude that the parallel-perpendicular transition in the TKLS model is discontinuous, similar to Fig.~\ref{fig:FORealOPSHZMain}. We used the order parameters $\Psi(\bm{k}_\parallel)$ and $\Psi(\bm{k}_\perp)$ to study this transition. In the parallel state, the separation between two types of particles is clearly observed, and $\Psi(\bm{k}_\parallel)$ reflects the charge fluctuation. On the other hand, with low density ($\rho<0.5$), two lanes formed by particles are typically separated by large stripes with low density  (Fig.~\ref{fig:ConfigurationTKLS}), and hence $\Phi(\bm{k}_\parallel)$ does not capture the transition. For the perpendicular state, both $\Psi(\bm{k}_\perp)$ and $\Phi(\bm{k}_\perp)$ show similar behaviour, we only show data for $\Psi(\bm{k}_\perp)$.

\subsection{Discussion : TKLS model}
\label{sec:discuss-tkls}

The TKLS model supports both parallel and perpendicular states as in the ELD model. However, unlike the ELD model, the parallel state exists in the TKLS even for low Peclet numbers. (The Peclet number of the TKLS model is less than unity, similar to the SHZ model.) Hence, this illustrates that large Peclet numbers are not necessary for the parallel state to emerge. We have focused on relatively high temperatures and large $E$ where non-equilibrium effects control these states; at lower temperatures and smaller fields, the phase separation would be affected by properties of the equilibrium phases, such as the surface tension between the phases of the Blume-Capel model.

Our interpretation of the resulting phenomenology is that the parallel and perpendicular states in the TKLS model have similar physical origins to the ELD model.  A traffic jam effect is responsible for the perpendicular state, recall Sec.~\ref{sec:eld-discuss}.  Compared with the SHZ model, the parallel state in the TKLS model is possible because of the attractive interactions between particles of the same charge and repulsive interactions between opposite charges.  In the ELD model, similar effective interaction arise from the dynamical rules, similar to the mechanism for MIPS in active matter~\cite{cates2015motility}. However, the two models' transitions between perpendicular and parallel states have different characters; the TKLS model has a hysteretic first-order transition, while the ELD transition takes place via an intermediate zig-zag state. 

We also remark that increasing $T$ in this model can drive a transition from the parallel to the perpendicular state.  This can be interpreted as an example of ``freezing by heating" \cite{helbing2000freezing}, where a low-current (frozen) state is stable at high temperature, while the low temperature (parallel) state is associated with a higher current.

Finally, two connections are notable between the TKLS and KLS models.  It is obvious that the TKLS model reduces to the KLS model if one removes all particles of a given color (say, blue), to obtain a single-species model.  However, it turns out that for total density $\rho=1$ (so there are no vacancies), one again recovers KLS behaviour where one color (say, blue) in the TKLS model plays the role of a vacancy in the KLS model.  In mapping between these cases, the values of $\rho,E$ have to be renormalised, and all hop rates are rescaled by $\gamma$ (because all moves in the TKLS description correspond to swaps). 

Since the KLS model only supports  disordered and parallel states (with no perpendicular state), this means that the TKLS phase diagram of Fig.~\ref{fig:PhaseDiagramTKLS} (obtained at density $\rho=0.5$) should have a significant dependence on density.  In fact, this is natural because the ``traffic jam'' interpretation of the perpendicular state requires a contrast between the dense region where motion is slow, and a dilute region where motion is fast.  If there are not enough vacancies to support a substantial dilute region, the traffic-jam instability will no longer operate.  One possibility is that the transition from disordered to perpendicular states exists for all densities, with a critical field $\mathcal{E}_*$ that diverges as $\rho\to1$, for example $\mathcal{E}_* \sim 1/(1-\rho)$.  Recalling that $E \simeq {\cal E}/(2L_\parallel)$, that would mean that the limits of large $L$ and $\rho\to1$ would not commute, so that $\rho=1$ is a singular point for the TKLS model.  However, this question is beyond the scope of this work.

\begin{table*}
\caption{\label{tab:ModelFeatures}Summary of features of the SHZ, ELD, and TKLS model.}
\begin{ruledtabular}
\begin{tabular}{llll}
 &SHZ Model&ELD Model&TKLS Model
\\ \hline
 Control Parameter&$E$,$\rho$,$\gamma$&$E$,$\rho$,$\gamma$,$\alpha$ &$E$,$\rho$,$\gamma$,$T$ \\
 Relevant Wavevector&$\bm{k}_{\perp}$&$\bm{k}_{\perp}$,$\bm{k}_{\parallel}$&$\bm{k}_{\perp}$,$\bm{k}_{\parallel}$\\
 \hline
 Enhanced Lateral Diffusion (ELD)&$\times$&\checkmark&$\times$\\
 Nearest-Neighbour Interaction&$\times$&$\times$&\checkmark\\
 \hline
 Disordered State& \checkmark&\checkmark&\checkmark \\
 Perpendicular State& \checkmark&\checkmark&\checkmark \\
 Parallel State& $\times$&\checkmark&\checkmark \\
\end{tabular}
\end{ruledtabular}
\end{table*}

\section{Conclusion and Outlook}
\label{sec:conclusion}

We have analysed three distinct models, which support different kinds of inhomogeneous states (Fig.~\ref{fig:SchematicPT}).
The basic properties of the models are summarised in Table \ref{tab:ModelFeatures}.  
This section summarises the implications of this work, the connections with the literature, and future directions.

For the SHZ model, the general behaviour and the form of the phase diagram were already established in previous work \cite{korniss1995novel, korniss1997nonequilibrium}.  
Still, the analysis in Figs.~\ref{fig:SORealOPSHZMain} and~\ref{fig:FORealOPSHZMain} characterises the associated phase transition in terms of its finite-size scaling, the distribution of its order parameter, and the spontaneous breaking of an $U(1)$ symmetry.  This analysis serves as a baseline for the other models considered, including that the transition from disordered to perpendicular state takes place at $E=O(1/L)$. We also showed that this model does not support a parallel state in contrast to the other models presented here.

The ELD model demonstrates that the parallel state is possible in a square lattice model.  This is achieved by a large hopping rate along the field (which enables large Peclet numbers) and by a strong enhancement of lateral diffusion when oppositely moving particles block each other.   Previous work~ \cite{dzubiella2002lane,klymko2016microscopic} suggested this state would not occur on square lattices with nearest-neighbour interaction.  We note, however, that the presence of the parallel state in the ELD model requires modelling assumptions that might be questionable in the colloidal context. In particular, lateral diffusion does not obey local detailed balance in the sense of~(\ref{equ:SHZ-detbal}).  Rather, the model serves as a proof-of-principle that such states can occur on the lattice in the presence of an effective interaction between particles of the same type caused by  the  enhanced lateral mobility.  
Our finite-size scaling analysis provides strong evidence that the transition between parallel and perpendicular states is a dynamical phase transition (and not, for example, a smooth crossover between two regimes).

As further evidence for this last hypothesis -- that an effective interaction leads to the parallel state -- we analysed the TKLS model, in which repulsion between the opposite types appears explicitly in the energy $H$.  This model indeed supports both parallel and perpendicular states.  However, the transition into the parallel state differs from the ELD case. In addition, we confirmed that the parallel state could also emerge for lower Peclet 
numbers. 
A similar transition between states with parallel and perpendicular ordering is also observed in two-species lattice models where all particles have attractive interactions (independent of species), and the two species have different concentrations~\cite{lyman2002phase, lyman2005steady}.

In the broader context, these results also raise a number of questions.  It is clear that models with particles hopping between lattice sites cannot be interpreted as quantitatively accurate models of colloidal particles moving in a solvent.  On the other hand, one might still hope that lattice models capture the dominant collective behaviour, including ``universal'' phenomena such as phase transitions.  The extent to which this holds is not clear in the current context.  For example, the perpendicular state has not been observed in simple Brownian dynamics simulations, although its existence is ``universal'' across the three lattice models considered here.  This point was also raised by~\cite{glanz2012nature,klymko2016microscopic}: transient ``traffic jams'' are observed in Brownian dynamics simulations, but they do not span the whole system, and there is no spontaneous breaking of translational symmetry.  In the comparison between on- and off-lattice models, a crucial aspect is that a single negatively charged (downwards-moving) particle on the lattice can block the upward motion of a large number of positively charged ones. This is not possible in Brownian dynamics, where a cluster of particles tends to drift with a velocity proportional to its net charge~\cite{klamser2021kinetic}.

Given the insights from this work about transitions into the parallel state and their analysis via order parameters and finite-size scaling, it would be interesting to revisit the behaviour of off-lattice (Brownian dynamics) models, to explore the similarities and differences with those found here.
Moreover, it is not even clear to what extent Brownian dynamics is a realistic description of a colloidal particle, where hydrodynamic interactions are also present~\cite{rex2008influence}.   This again raises the question of the extent to which simplified models can capture universal properties of these non-equilibrium systems.

We end with a brief discussion of two issues that have been raised by previous work. First, the behaviour of this system depends on the shape of the simulation box, particularly the aspect ratio $L_\perp/L_\parallel$, which has been set to unity in this work. 
It is expected that long ($L_{\parallel} \gg L_{\perp}$) systems prefer perpendicular states as a small fraction of particles can create traffic jams and block the entire system. On the other hand, in wide ($L_{\parallel} \ll L_{\perp}$) systems, large fraction of particles are needed to achieve jamming. In addition, when laning/parallel separation does happen, wide systems prefer multiple lanes and non-equilibrium steady states are aspect ratio-dependent~\cite{shaw1999phase, zia2000possible, levine2001ordering}. 

Our results for these lattice models are consistent with a dynamical phase separation, as evidenced by the order parameters $\Phi$ and $\Psi$ having non-zero limits as we increase the system size and approach the limit $L\to\infty$ (keeping always $L_\parallel=L_\perp$). 
 It is known for other lattice models~\cite{levine2001ordering} that non-equilibrium steady states may be system shape dependent, so different results might be obtained for different aspect ratios.
 
 In addition, comparing with Brownian dynamics studies, Klymko et. al.~\cite{klymko2016microscopic} found that the parallel state is stable: starting from that state, opposite types of particles never mix throughout their simulations. However, Glanz and L{\"o}wen~\cite{glanz2012nature} claimed the correlation length of the system is finite in the thermodynamic limit and hence that the system does not exhibit a phase transition.  More recent results~\cite{poncet2017universal} also indicate there is no phase separation associated with laning, but that correlations are algebraic.
 
 Given the differences between lattice and off-lattice models, our results do not speak directly to the existence of phase transitions in Brownian dynamics, but the results presented here do indicate the possibility of parallel phase separation induced by ELD, as argued in~\cite{klymko2016microscopic}.  
It might be that different criteria for identifying phase transitions are  yielding different conclusions in~\cite{klymko2016microscopic,glanz2012nature,poncet2017universal}: in the absence of a free energy, it would be useful to settle on the most appropriate criterion for establishing (or disproving) the existence of such transitions.

A second open question is the process by which the system evolves into its steady state.  So far, the studies of two-species driven systems are mainly focused on steady state behaviours, and the time-evolution of the system is little known in two-dimension. For the approach to the steady state, we have presented illustrative results in Figs.~\ref{fig:ConfigurationSHZ}, \ref{fig:ConfigurationMIPS}, \ref{fig:ConfigurationTKLS}. However, all quantitative data were obtained in the steady state.  A detailed analysis of the time-dependent behaviour would be useful, especially given that some surprising results were already obtained for the SHZ model~\cite{kafri2000slow,mettetal2002coarsening,georgiev2005anomalous} and the KLS model~\cite{yeung1992phase, alexander1996monte, levine2001ordering}.

\begin{acknowledgments}
We thank Daan Frenkel, Tal Agranov, Katie Klymko, Maria Bruna, and Mike Cates for discussions.
KT is grateful to the (EPSRC) for funding through grant EP/T031247/1.
\end{acknowledgments}

\appendix

\section{MC dynamics}
\label{sec:app-mc}

\subsection{SHZ and TKLS models}\label{sec:app-shz}
In Monte-Carlo (MC) simulations, the transition rate $W$ between two configurations can be decomposed into two parts, the proposal rate $w_{\rm prop}$ and the acceptance probability $p_{\rm acc}$:
\begin{equation}
  W = w_{\rm prop} \times p_{\rm acc} \;.
\end{equation}
Hence, we have the liberty to choose the proposal rate and acceptance rate as long as their product remains the same, and we exploit this to improve the efficiency of the simulations.
In practice, the proposal rate is further decomposed as $w_{\rm prop} = w_0 p_{\rm prop}$ where $w_0$ is the total rate of MC updates, and $p_{\rm prop}$ is the probability of proposing the specific update of interest.  In the continuous-time MC approach~\cite{newman1999monte}, this means that for each MC update, the simulation time is updated by an exponentially distributed random variable of mean $(1/w_0)$.

In a simple MC simulation~\cite{metropolis1953equation}, a uniform proposal rate is sufficient.  For example, 
in the SHZ and TKLS models, the proposal probability is simply $p_{\rm prop} = 1/(4N)$, which corresponds to first picking a random particle ($1/N$) and then a random direction ($1/4$). 
We take $w_0=4N$ so that all moves are proposed with unit rate.  Then
the acceptance probabilities are simply the values of the rates, as given in Eq.~\ref{eq:SHZRegularHopRate}, \ref{eq:SHZRegularSwapRate} and Eq.~\ref{eq:TKLSphRate}, \ref{eq:TKLSppRate}. 
It is necessary for this approach that the values of all rates are less than unity.

The dynamics of the SHZ and TKLS model are then:

\noindent
(i) Initialise the system with a random configuration of $N$ particles and start simulation at $t=0$.

\noindent
(ii) Increase the time $t$ by an exponential random variable with mean $(1/w_0)$, where $w_0=4 N$, as above.

\noindent
(iii) Select a random particle with probability $1/N$ and select a random direction with probability $1/4$.

\noindent
(iv) Accept the move with a probability that is equal to the appropriate rate. (For SHZ, this is either Eq.~\ref{eq:SHZRegularHopRate} or \ref{eq:SHZRegularSwapRate}, depending on whether the target site is empty or occupied). 

\noindent
(v) Repeat steps (ii)-(iv) until $t\geq t_{\text{max}}$ and terminate the simulation. 

\subsection{ELD model}\label{sec:app-eld-algo}
In contrast to the SHZ and TKLS models described above, the uniform proposal rate is not appropriate for ELD model, due to the large differences in rates between the blocked and unblocked cases. Instead, we propose moves based on the exponential rates given in Eq.~\ref{eq:MIPSRegularHopRate},\ref{eq:MIPSSpecialHopRate}. 

As described in Sec~\ref{sec:eld-def}, we have four sets of dynamical rules, i.e. regular hop, regular swap, blocked hop, and blocked swap. The dynamics are implemented in the following way.  For a given particle at site $\bm{x}$, the regular hop rates are given by $W_{\text{ELD,RH}}(\bm{x}, \bm{e}, \sigma_{\bm{x}})$, with $\bm{e}=(1, 0)$, $(-1, 0)$, $(0, 1)$, $(0, -1)$ for right, left, forward, backward regular hops respectively. And the blocked hop rates are given by $W_{\text{ELD,BH}}(\bm{x}, \bm{e}, \sigma_{\bm{x}})$, as in Eq.~\ref{eq:MIPSRegularHopRate},\ref{eq:MIPSSpecialHopRate}.

We first calculate the escape rate of the regular hop and blocked hop:
\begin{equation}
\begin{aligned}
  \mu_{\text{ELD, RH}} & = \sum_{\bm{e}}
  W_{\text{ELD,RH}}(\bm{x}, \bm{e}, \sigma_{\bm{x}}),
  \\
    \mu_{\text{ELD, BH}} & = \sum_{\bm{e}}
  W_{\text{ELD,BH}}(\bm{x}, \bm{e}, \sigma_{\bm{x}}).
\end{aligned}
\end{equation}
Then define
\begin{equation}
  \mu = \max{(\mu_{\text{ELD, RH}}, \mu_{\text{ELD, BH}})}.
\end{equation}
The total rate of MC updates will be $w_0=\mu N$.
We also define the hop proposal probabilities
\begin{equation}
\begin{aligned}
p_{\text{prop,RH}}(\bm{x}, \bm{e}, \sigma_{\bm{x}})
&=\frac{W_{\text{ELD,RH}}(\bm{x}, \bm{e},
\sigma_{\bm{x}})}{\mu}
\\
p_{\text{prop,BH}}(\bm{x}, \bm{e}, \sigma_{\bm{x}})
&=\frac{W_{\text{ELD,BH}}(\bm{x}, \bm{e},
\sigma_{\bm{x}})}{\mu}
\end{aligned}
\label{equ:propELD}
\end{equation}

The resulting simulation procedure is:

\noindent
(i) Initialise the system with a random configuration of $N$ particles and start simulation at $t=0$.

\noindent
(ii) Increase the time $t$ by an exponential random variable with mean $\frac{1}{\mu N}$, where $N$ is number of particles in the system. 

\noindent
(iii) Select a random particle with probability $1/N$, and check if the selected particle is blocked in its forward direction by a particle of the opposite type.
    
(a) If the particle is not blocked, we propose a move in direction $\bm{e}$ with regular proposal probability $p_{\text{prop,RH}}(\bm{x}, \bm{e}, \sigma_{\bm{x}})$.
    
(b) If the proposed particle is blocked, we propose a move in direction $\bm{e}$ with blocked proposal probability $p_{\text{prop,BH}}(\bm{x}, \bm{e}, \sigma_{\bm{x}})$.

\noindent
(iv) Accept the move with probability $1$ or $\gamma$ depending on whether the move is a particle hop or particle swap.  We note for blocked particle, the blocked forward hop is forbidden.

\noindent
(v) Repeat steps (ii)-(iv) until $t\geq t_{\text{max}}$ and terminate the simulation.

Note: The definition of $\mu$ together with \eqref{equ:propELD} ensures that the larger of $\sum_{\bm{e}} p_{\text{prop,RH}}$ and $\sum_{\bm{e}} p_{\text{prop,BH}}$ is equal to unity.  In steps (iii)(a,b), there are always four choices for $\bm{e}$, but the sum of the probabilities for the four cases may be less than unity.  In this case, there is a finite probability that no move is proposed at all, and step (iv) is skipped. 

\section{Linear Stability Calculation of ELD model}\label{sec:Stability_analysis}

\subsection{Perpendicular Instability}\label{sec:Stability_analysis_perp}
In this section, we give the derivation of the linear instabilities presented in the ELD model (Sec~\ref{sec:eld-linear}).
A simple theory for the SHZ model is given by Eq.~\ref{eq:HydrodynamicEq}.   As in \eqref{eq:Perturbation}, we consider a small perturbation $\delta \rho = (\delta \rho_+, \delta \rho_-)$, so that Eq.~\ref{eq:HydrodynamicEq} can be linearised.  It and takes the form $\delta \dot{\rho}_{\bm{k}} = \mathcal{L}_{\bm{k}}\delta \rho_{\bm{k}}$, where $\mathcal{L}_{\bm{k}}$ is a matrix that contains dynamical information.  The exponential decay rate of the perturbation is $\lambda$, which obeys the eigenvalue equation
%We can consider a perturbation about an homogeneous state of the form given by Eq.~\ref{eq:Perturbation}. Linearisation of Eq.~\ref{eq:HydrodynamicEq} around the constant density profile of two species of particles can be written as 
\begin{equation}\label{eq:app-linearisation}
    -\lambda A (\bm{k})
    =\mathcal{L}_{\bm{k}} A (\bm{k}),
\end{equation}
where $A (\bm{k}) = (A_+ (\bm{k}), A_- (\bm{k}))$ is the amplitude of the perturbation and
\begin{align}
    \mathcal{L}_{\bm{k}}=
    \begin{pmatrix}
        -D\lVert\bm{k}\rVert^2-\epsilon(1-\frac{3}{2}\rho)i k_{\perp}&
        \epsilon \frac{\rho}{2} i k_{\perp}\\
        -\epsilon \frac{\rho}{2} i k_{\perp}&
        -D\lVert\bm{k}\rVert^2+\epsilon(1-\frac{3}{2}\rho)i k_{\perp}
    \end{pmatrix},
\end{align}
where $\lVert\bm{k}\rVert^2=k_{\perp}^2+k_{\parallel}^2$ is the norm of wavevector $\bm{k}$.
The two eigenvalues of the matrix $-\mathcal{L}_{\bm{k}}$ are  
\begin{equation}%\label{eq:eigenvalue}
    \lambda^{\pm}_{\bm{k}}=D\lVert\bm{k}\rVert^2\pm\epsilon k_{\perp}\sqrt{(1-\rho)(2\rho-1)},
\end{equation}
and the condition for the homogeneous solution to be unstable is that the lowest eigenvalue is negative.  
The onset of the instability is given by the smallest wavevectors: in this case, $\bm{k}=\bm{k}_\perp$
The instability only occurs if $\rho>(1/2)$, in which case the system is unstable for
\beq
   \epsilon > \epsilon^* =\frac{2\pi D}{L_{\parallel}}\sqrt{\frac{1}{(1-\rho)(2\rho-1)}}.
\eeq
which is (\ref{equ:eps-star}) of the main text.
With the eigenvalues we calculated above, we can also obtain the eigenvectors when the instabilty happens. Let $\bm{v}=(v_+, v_-)$ be the eigenvector so we have,
\beq
   \mathcal{L}_{\bm{k}}\bm{v} = -\lambda^{-}_{\bm{k}}\bm{v},
\eeq
as $\lambda^{+}$ is always stable. The eigenvector $\bm{v}$ is 
\begin{align}
    \bm{v}=
    \begin{pmatrix}
        1\\
        e^{-i\theta}
    \end{pmatrix},
\end{align}
 where $\theta = \arccos{((2/\rho)-3)}$ and $0\leq \theta\leq \pi$.  The instability occurs via growth of regions where either $\rho_+$ or $\rho_-$ is large, and the relative phase $\theta$ determines the distance between points where these two densities are maximal. For small $\rho$ [close to $(1/2)$] then the two dense regions have positions close to each other (small $\theta$) while for $\rho\approx1$ then they are far apart ($\theta\approx \pi$).
 
\subsection{Parallel Instability}\label{sec:Stability_analysis_para}
As discussed in Sec.~\ref{sec:eld-linear}, the ELD model has an environment dependent diffusion in $x$ direction and this effect may be captured by an anisotropic theory similar to MIPS \cite{cates2015motility} given by Eq.~\ref{eq:HydrodynamicEqMIPS}.
In that equation, $D_{yy}$ is a constant diffusivity in the $y$-direction, while the diffusivity $D_{xx}$ in the $x$-direction depends on the concentration of the of the other species.  For the parallel instability, we can linearise the hydrodynamic equation again and obtain a form similar to Eq.~\ref{eq:app-linearisation} with matrix 
\begin{align}
    \mathcal{L}_{\bm{k}}=
    \begin{pmatrix}
        -k^{2}_{\perp}D_{xx}(\rho_-)&
        -\frac{\rho}{2} D^{'}_{xx}(\rho_-) k^2_{\perp}\\
        -\frac{\rho}{2} D^{'}_{xx}(\rho_+) k^2_{\perp}&
        -k^{2}_{\perp}D_{xx}(\rho_+)
    \end{pmatrix}.
\end{align}
where we restricted to $\bm{k}=(k_\perp,0)$ to simplify our calculation. (The parallel component does not affect the instability to laning.)
Again, the condition for instability is given by negative eigenvalue: it takes place when
\begin{equation}
    \frac{2}{\rho}<\frac{|D^{\prime}_{xx}(\rho)|}{D_{xx}(\rho)},
\end{equation}
which is (\ref{equ:mips-instability}) of the main text.
This analysis suggests that the instabilities do not depend on the wavevector, and instabilities occur at all scales, consistent with numerical simulation.

We can also obtain the eigenvectors $\bm{v}=(v_+, v_-)$ associated with the eigenvalues to be $\bm{v}=(1, -1)$ and $\bm{v}=(1, 1)$.  The case $\bm{v}=(1, 1)$ corresponds to $D^{\prime}_{xx}(\rho)<0$ and particles form a single cluster, which does not occur in our system.  The case $\bm{v}=(1, -1)$ corresponds to $D^{\prime}_{xx}(\rho)>0$. Physically, it corresponds to the laning effect, where the two types of particles phase separate, parallel to the external field.

\nocite{*}

\bibliography{apssamp}% Produces the bibliography via BibTeX.

\end{document}